\documentclass[final, 5p, times, twocolumn]{elsarticle}
 
\usepackage{amsmath}  
\usepackage{amssymb}
\usepackage{algorithm}
\usepackage{makecell}
\usepackage{graphicx}
\usepackage{balance}
\usepackage{subfig}
\usepackage{comment}
\usepackage{algorithmic}
\usepackage[utf8]{inputenc}
\usepackage{kotex}
\usepackage{tabularx}
\usepackage{booktabs}
\usepackage{multirow}
\usepackage{threeparttable}

\usepackage{xurl}
\usepackage[hidelinks]{hyperref}
\makeatletter
\g@addto@macro{\UrlBreaks}{\do\/\do\-}
\makeatother

\usepackage{natbib}
\biboptions{numbers}
\usepackage{etoolbox}
\usepackage{microtype}
\usepackage{doi}

\usepackage{kotex}

\begin{document}

\begin{frontmatter}

\title{ShuntServe: Cost-Efficient LLM Serving on Heterogeneous Spot GPU Clusters}

\author[hyu-ds]{Seungwoo Jeong}
\author[hyu-ai]{Moohyun Song}
\author[hyu-ds]{Juhyun Park}
\author[hyu-ds]{Kyungyong Lee\corref{cor1}}
\ead{kyungyong@hanyang.ac.kr}
\cortext[cor1]{Corresponding author}
\affiliation[hyu-ds]{organization={Hanyang University},
            addressline={Department of Data Science}, 
            city={Seoul},
            postcode={04763}, 
            country={Republic of Korea}}
\affiliation[hyu-ai]{organization={Hanyang University},
            addressline={Department of Artificial Intelligence}, 
            city={Seoul},
            postcode={04763}, 
            country={Republic of Korea}}

\begin{abstract}
As large language model (LLM) services become widely adopted, the cost of GPU resources for serving these models in cloud environments has emerged as a critical concern. Spot instances offer up to 90\% cost savings over on-demand instances, but their frequent interruptions and limited availability pose significant challenges for continuous LLM serving. GPU spot instances, in particular, exhibit lower and more volatile availability than CPU-based instances, making homogeneous clusters that depend on a single GPU type vulnerable to correlated failures. Heterogeneous clusters spanning multiple GPU types can address this by leveraging complementary availability patterns across diverse spot pools, yet existing LLM serving systems are designed for homogeneous environments and suffer from load imbalance when deployed on heterogeneous GPUs. This paper presents ShuntServe, a cost-efficient LLM serving system for heterogeneous spot GPU clusters. ShuntServe employs a roofline model-based analytical serving performance estimator and a dynamic programming-based model placement optimizer that jointly determines node configuration, parallelization strategy, and layer assignment to maximize throughput across heterogeneous GPUs. To enhance fault tolerance when using spot instances, ShuntServe combines output-preserving request migration with concurrent initialization via a shared tensor store, minimizing migration downtime by overlapping replacement node preparation with ongoing serving. Evaluation on Llama-3.1-70B and Qwen3-32B with a heterogeneous AWS cluster of L4, A10G, and L40S GPUs shows that ShuntServe achieves 1.42$\times$ and 1.35$\times$ higher throughput than state-of-the-art baselines and attains 31.9\% and 31.2\% cost efficiency improvements over on-demand instances for offline and online serving, respectively.
\end{abstract}

\end{frontmatter}

\section{Introduction}
Large language models (LLMs) become a fundamental component for a broad spectrum of applications, from conversational assistants and code generation to domain-specific reasoning~\cite{grattafiori2024llama3herdmodels}. While commercial LLM APIs are widely available, many organizations choose to deploy models independently due to prohibitive per-query costs at scale, data privacy and regulatory requirements, or the need for specialized fine-tuned models that public services cannot accommodate.

Running self-hosting LLM services, however, can become costly. LLM inference is GPU-intensive, and demand can fluctuate significantly, making on-premise clusters with large upfront capital expenditures difficult to justify. Cloud computing offers elastic GPU provisioning as an alternative, yet GPU instances are priced significantly higher than their CPU-only counterparts. For example, serving large models such as Llama-3-70B or Llama-3-405B on clusters of high-end GPUs can cost tens of thousands of dollars per month.

Spot instances can reduce these costs by up to 90\% by leveraging providers' idle capacity~\cite{spot-instance-analysis, spot-analysis-javadi}, but they may be reclaimed with only a short grace period, threatening service continuity. This risk is especially pronounced for GPU workloads, where empirical studies show that GPU spot instances exhibit significantly lower and more volatile availability than CPU-based instances~\cite{multinode-spot-dataset, interrupt-visible-www, spotlake-iiswc, multi-spotlake-www}. Since spot instances are provisioned from the provider's surplus capacity for each specific instance type, their availability is inherently tied to the demand fluctuations of that type. Thus, relying on a single GPU type for a homogeneous cluster is inherently fragile, as correlated shortages can simultaneously disable all instances of that type. Heterogeneous clusters spanning multiple GPU types can instead exploit complementary availability patterns across diverse spot pools, improving both resilience and cost efficiency.

However, existing serving systems are mostly designed for homogeneous GPU environments, assuming identical model partitions and uniform workload distribution. In heterogeneous clusters, GPUs with disparate compute capabilities and memory bandwidths receive equal workloads, causing slower devices to bottleneck the pipeline. Moreover, the distinct characteristics of LLM inference, compute-bound prefill and memory-bound decode~\cite{distserve,splitwise,thunderserve,pmlr-v267-jiang25c}, create decoupled bottlenecks across heterogeneous stages. This necessitates a serving system that jointly optimizes model placement, parallelization strategy, and layer assignment based on each device's unique capabilities.

To this end, we propose ShuntServe, a cost-efficient LLM serving system designed for heterogeneous spot GPU clusters. ShuntServe combines a roofline model-based analytical serving performance estimator with a dynamic programming (DP) algorithm to determine throughput-maximizing model placements that account for the distinct compute and memory characteristics of different GPU types. To handle spot GPU instance interruptions, ShuntServe employs output-preserving request migration, which recovers in-flight requests by recomputating their generated outputs on replacement instances rather than transferring state from the interrupted instances. ShuntServe further minimizes migration downtime through concurrent initialization, which decouples the inference engine lifecycle from model weights loading via a shared tensor store. This allows the replacement instance's model loading and engine initialization to proceed in parallel with ongoing serving on the existing pipeline, significantly reducing the time required to resume full pipeline operation after an interruption.

We evaluate ShuntServe using Llama-3.1-70B and Qwen3-32B on a heterogeneous AWS GPU cluster comprising L4, A10G, and L40S instances with realistic workload traces. Through optimized heterogeneous model placement, ShuntServe achieves up to 1.42$\times$ higher throughput than the state-of-the-art homogeneous serving system and up to 1.35$\times$ higher throughput than state-of-the-art heterogeneous baselines, without sacrificing request latency. By leveraging spot instances with its fault tolerance mechanisms, ShuntServe attains 31.9\% and 31.2\% cost efficiency improvements over on-demand instances for offline and online serving, respectively.

The key contributions of this paper are as follows:
\begin{itemize}
    \item \textbf{Analytical serving performance estimator.} An analytical roofline model-based approach that estimates serving throughput across diverse parallelization configurations and heterogeneous GPU types without comprehensive profiling.
    
    \item \textbf{DP-based model placement optimizer.} A DP algorithm augmented with beam search that jointly optimizes node configuration, parallelization strategy, and layer assignment to maximize throughput in heterogeneous clusters.
    
    \item \textbf{Fault tolerance for spot interruptions.} recomputation-based output-preserving request migration combined with concurrent initialization via a shared tensor store, minimizing migration downtime and maximizing spot utilization.
\end{itemize}

\section{LLM Serving with Heterogeneous GPUs}
\label{sec:distributed_llm_serving_heterogeneous_gpus}
The scale of modern LLMs has reached tens to hundreds of billions of parameters, rendering single-GPU serving infeasible \cite{kaplan2020scalinglawsneurallanguage, henighan2020scalinglawsautoregressivegenerative}. This challenge is typically addressed using two primary parallelization strategies of \emph{Tensor Parallelism (TP)} and \emph{Pipeline Parallelism (PP)}. TP partitions operators within each layer to run in parallel, which reduces request latency but is highly network-sensitive due to frequent collective communications (e.g., AllReduce) \cite{megatron_LM}. In contrast, PP divides the model into stages, groups of layers, that are assigned to different devices \cite{gpipe}. While this does not reduce single-request latency, it increases overall throughput by processing micro-batches in parallel and is more broadly applicable under constrained network environments, as communication only occurs at stage boundaries.

Serving large-scale LLMs requires many GPUs that incurs significant operational costs. This economic constraint renders performance-only optimization insufficient, and new approaches are necessary to co-optimize for high throughput and cost-effectiveness.

\subsection{Cost Savings and Vulnerability of Spot Instances}
\label{subsec:spot_serving_challenges}

Major cloud providers offer spot instances at discounts of up to 90\% by utilizing idle compute resources \cite{spot-instance-analysis, spot-analysis-javadi}. However, these instances are subject to interruption that a provider reclaims the instance after a short grace period when resources become scarce.

Because of the volatility, spot instances are suitable for stateless workloads, but using them for distributed LLM serving is challenging. Unlike traditional single-node inference that could often finish serving tasks within the grace period \cite{spot-for-online-service, spotweb}, LLM serving is quasi-stateful that requires maintaining a persistent KV cache for long-running, autoregressive requests. In a distributed setting, this state creates a strong inter-node dependency.

Consequently, this stateful and distributed nature renders a simple finish and migrate strategy on a small-scale serving environment impractical. The interruption of a single node triggers a cascading failure that stalls the entire pipeline. This inherent fragility makes leveraging spot instances exceptionally difficult, mandating an effective fault tolerance mechanism.

\begin{figure}[t]
\centering
\includegraphics[width=0.98\columnwidth]{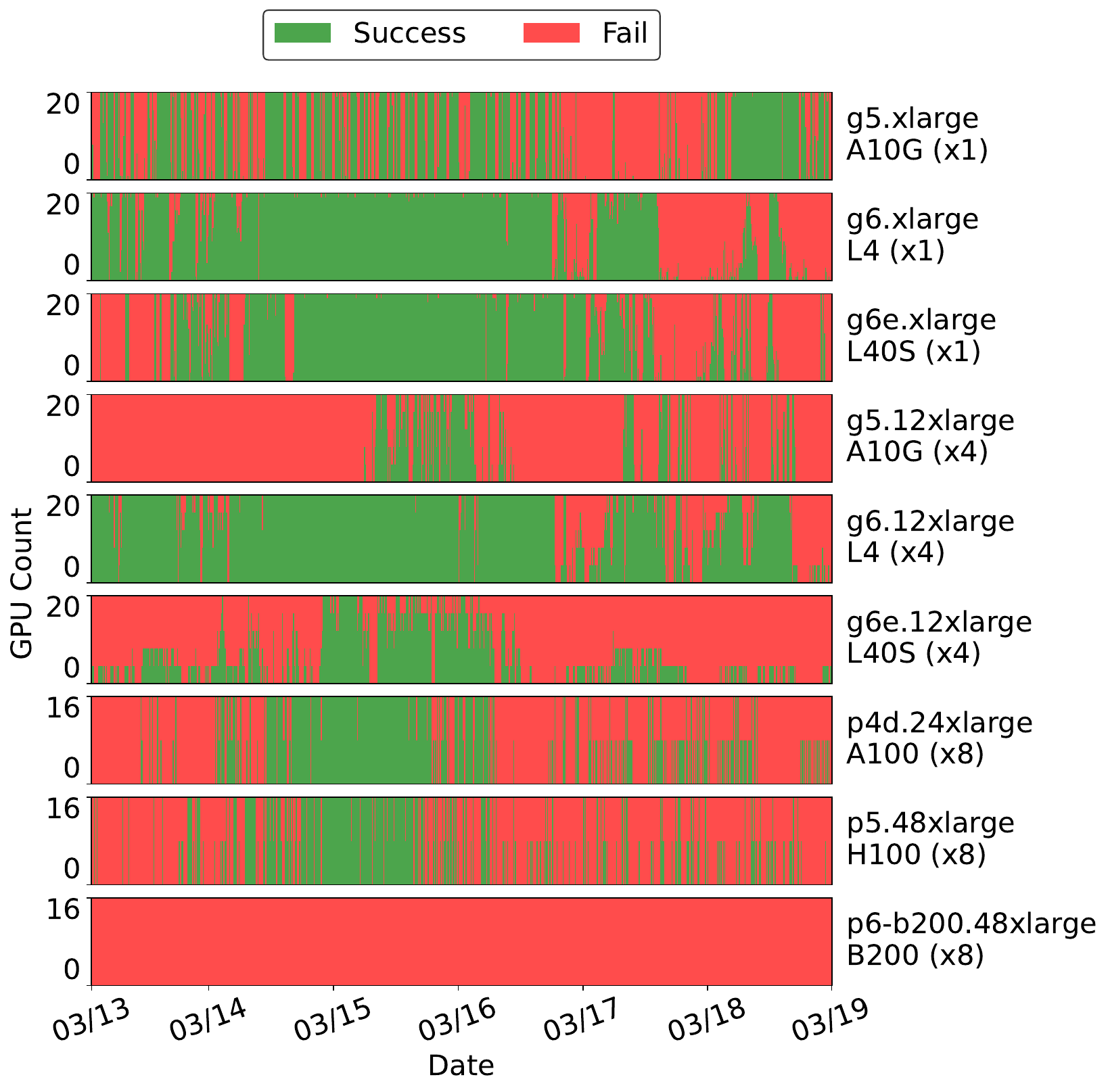}
\caption{Time-series Graph of Available GPU Instance Counts by Type in AWS (us-west-2).}
\label{fig:spot_availability}
\end{figure}

\begin{figure*}[t]
\centering
\includegraphics[width=0.97\linewidth]{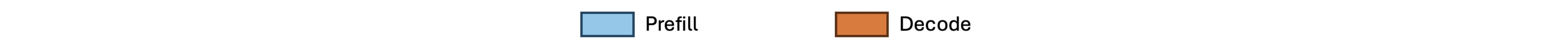}
\subfloat[Even partitioning.\label{fig:hetero_gpu_even_partitioning}]{%
    \includegraphics[width=0.32\textwidth]{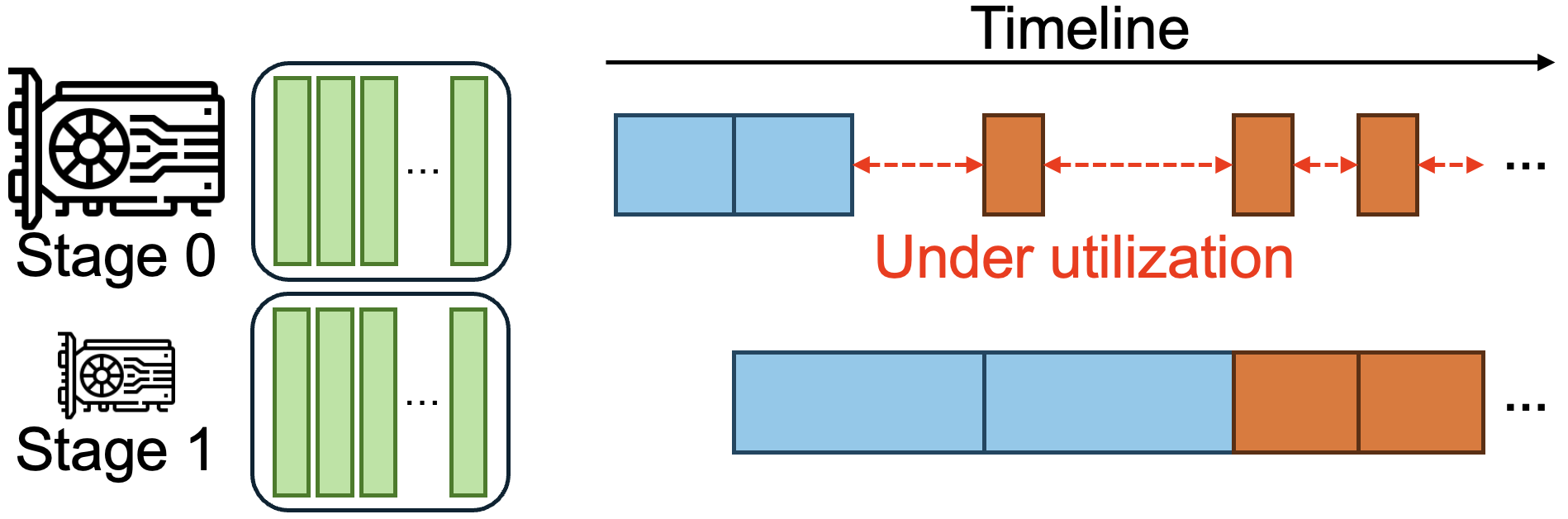}}
\hfill
\subfloat[Uneven layer partitioning.\label{fig:hetero_gpu_uneven_partitioning}]{%
    \includegraphics[width=0.32\textwidth]{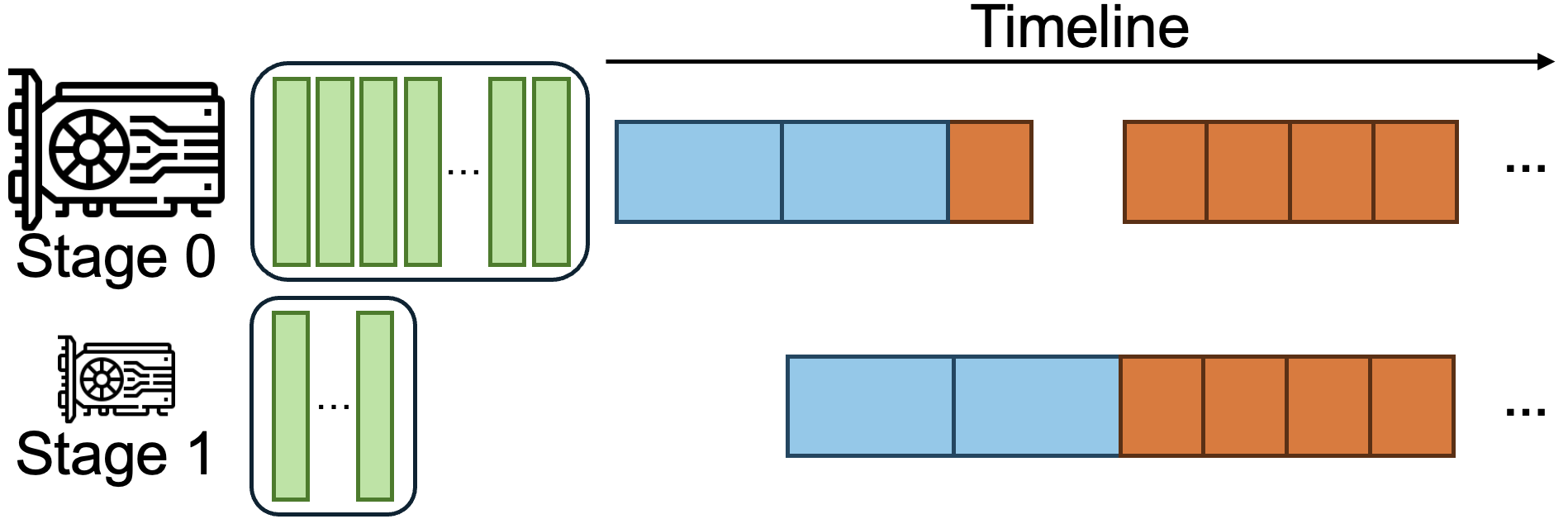}}
\hfill
\subfloat[Per-stage TP.\label{fig:hetero_gpu_per_stage_partitioning}]{%
    \includegraphics[width=0.32\textwidth]{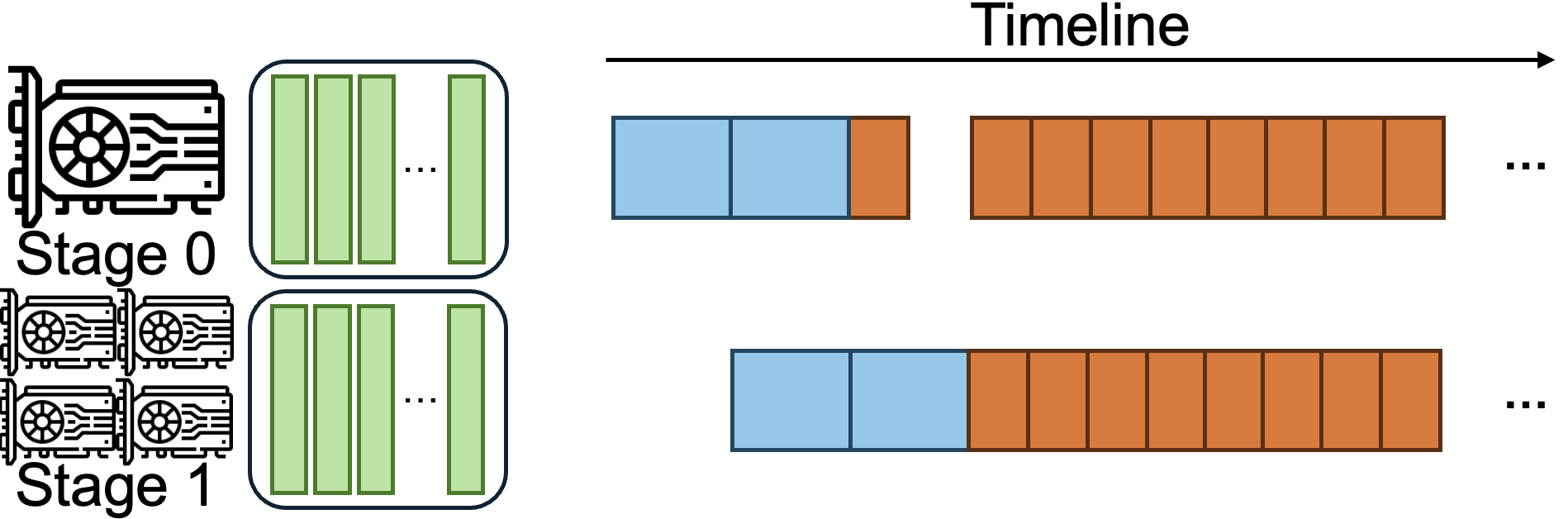}}
\caption{Asymmetric model partitioning methods to mitigate performance bottlenecks in heterogeneous GPU clusters.}
\label{fig:hetero_per_stage_partitioning}
\end{figure*}

\subsection{Spot GPU Scarcity and Necessity of Heterogeneity}
\label{subsec:availability_and_heterogeneity}
The spot availability of GPU instances is markedly lower than that of CPU-based instances and varies sharply across GPU types, with high-end accelerators particularly scarce~\cite{multinode-spot-dataset, interrupt-visible-www, spotlake-iiswc, multi-spotlake-www}. Figure~\ref{fig:spot_availability} illustrates the hourly spot availability of GPU instance types in the AWS us-west-2 region from March 13 to 18, 2026 (UTC) where the y-axis represents the number of GPUs and the x-axis shows the timeline, high-end GPU instances are frequently unavailable that is consistent with prior reports on GPU scarcity \cite{helix}. Our 6-day monitoring revealed that H100 GPUs adequate for serving Llama-3-405B were available only 28.64\% of the time, and even B200 GPUs were never successfully provisioned throughout the entire observation period. In contrast, Mid-tier instances (i.e., g5, g6, g6e) exhibit different and often more stable availability patterns.

This high volatility and scarcity pose a significant operational challenge for systems relying on homogeneous clusters. Attempting to build a cluster from a single, high-end GPU type would nullify potential cost savings, as those instances are rarely available.

Heterogeneous GPU clusters are an effective solution to this limitation. Different instance types often exhibit complementary availability patterns; as shown in Figure~\ref{fig:spot_availability}, g5/g6/g6e instances are available when p4d/p5/p6 instances are not. This approach enables cost savings by maximizing spot instance utilization from diverse pools, minimizing reliance on expensive on-demand instances.

\begin{table}[t]
\centering
\caption{Comparison of hardware specifications for various GPUs. FLOPS values are based on BF16 non-sparse operations.}
\small  
\label{tab:heterogeneous_gpus_spec}
\begin{tabular}{l|c|c|c}
\toprule
GPU & \begin{tabular}[c]{@{}c@{}}Mem. (GB)\end{tabular} & \begin{tabular}[c]{@{}c@{}}FLOPS\\(TFLOPS)\end{tabular} & \begin{tabular}[c]{@{}c@{}}Mem. BW. (GB/s)\end{tabular} \\ 
\hline
L4 & 24 & 121 & 300 \\
A10G & 24 & 70  & 600 \\
L40S & 48 & 362 & 864 \\ 
A100 & 40 & 312 & 1,555 \\ 
H100 & 80 & 989 & 3,350 \\ 
B200 & 180 & 4,500 & 7,700 \\
\bottomrule
\end{tabular}
\end{table}

\subsection{Asymmetric Model Partitioning}
\label{subsec:asymmetric_model_partitioning}

Deploying distributed LLM serving on a heterogeneous GPU cluster introduces a new challenge for model partitioning. This arises from the disparities in floating-point operations per second (FLOPS), which represents compute capability, and memory bandwidth (Mem. BW.) across different GPU types, as shown in Table~\ref{tab:heterogeneous_gpus_spec}.

\paragraph{Uneven Layer Partitioning in PP}
In homogeneous environments, a balanced partitioning scheme that assigns an equal number of layers per GPU is standard. However, applying this naive approach to a heterogeneous setting causes the least powerful GPU to become a performance bottleneck, severely limiting pipeline throughput. Figure~\ref{fig:hetero_gpu_even_partitioning} illustrates that the high-performance GPU (Stage 0) is forced to be idle while waiting for the low-performance GPU (Stage 1). To resolve this, layers must be partitioned asymmetrically, proportional to GPU's processing capability, to balance the execution time of each stage, reducing idle time and improving pipeline throughput (Figure~\ref{fig:hetero_gpu_uneven_partitioning}).

\paragraph{Per-Stage TP}
While uneven layer partitioning balances stages to improve throughput, the end-to-end request latency, which is the sum of all stage times, may remain high due to the inclusion of low-performance GPUs. A complementary approach is to configure multiple low-performance GPUs as a single stage using TP, thereby increasing the processing speed of a stage. For instance, as shown in Figure~\ref{fig:hetero_gpu_per_stage_partitioning}, configuring four low-performance GPUs with a TP degree of 4 achieves comparable performance to that of a single high-performance GPU in Stage 0. This allows stages composed of weaker GPUs to balance their execution time against stages with high-performance GPUs, reducing pipeline bottlenecks and increasing overall throughput~\cite{helix,hexgen}.

\begin{figure}[t]
    \centering
    \includegraphics[width=\linewidth]{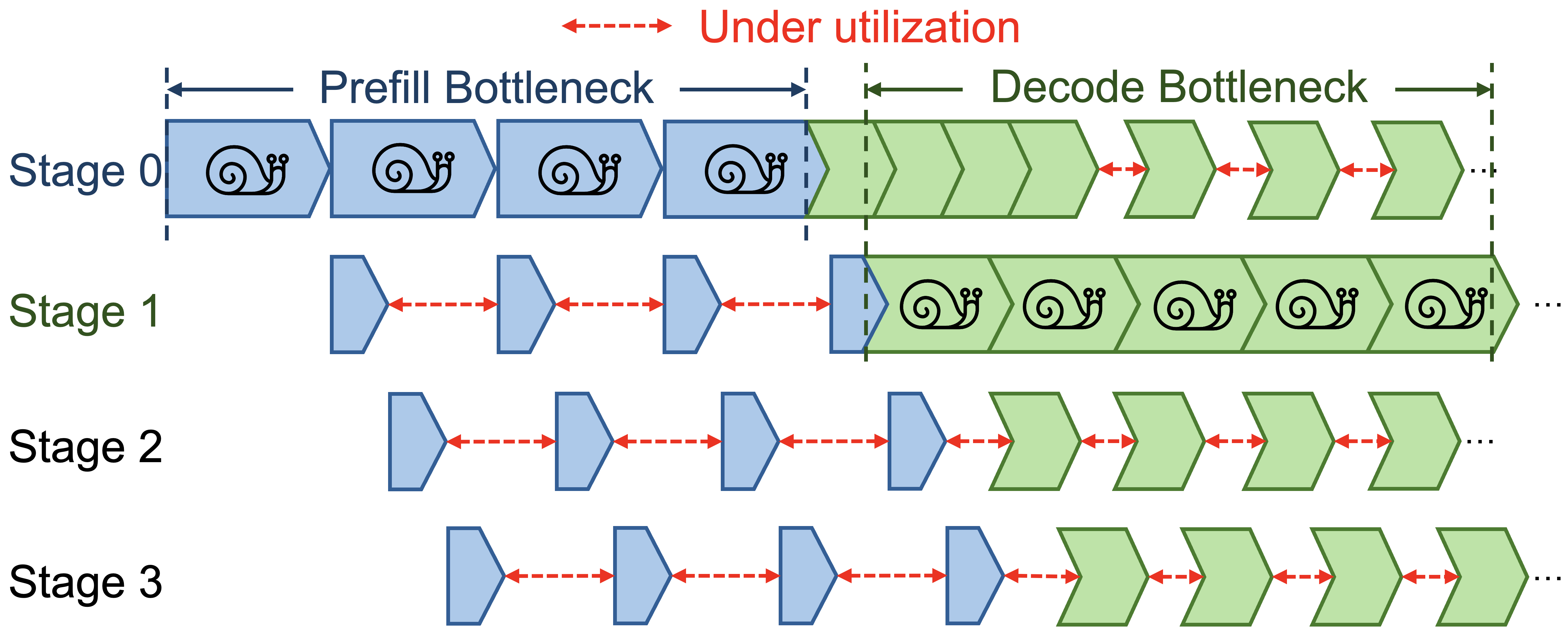}
    \caption{Decoupled bottlenecks in the prefill and decode phase on a pipeline with heterogeneous capability GPUs.}
    \label{fig:decoupled_bottleneck}
\end{figure}

\begin{figure}[t]
\centering
\includegraphics[width=0.45\textwidth]{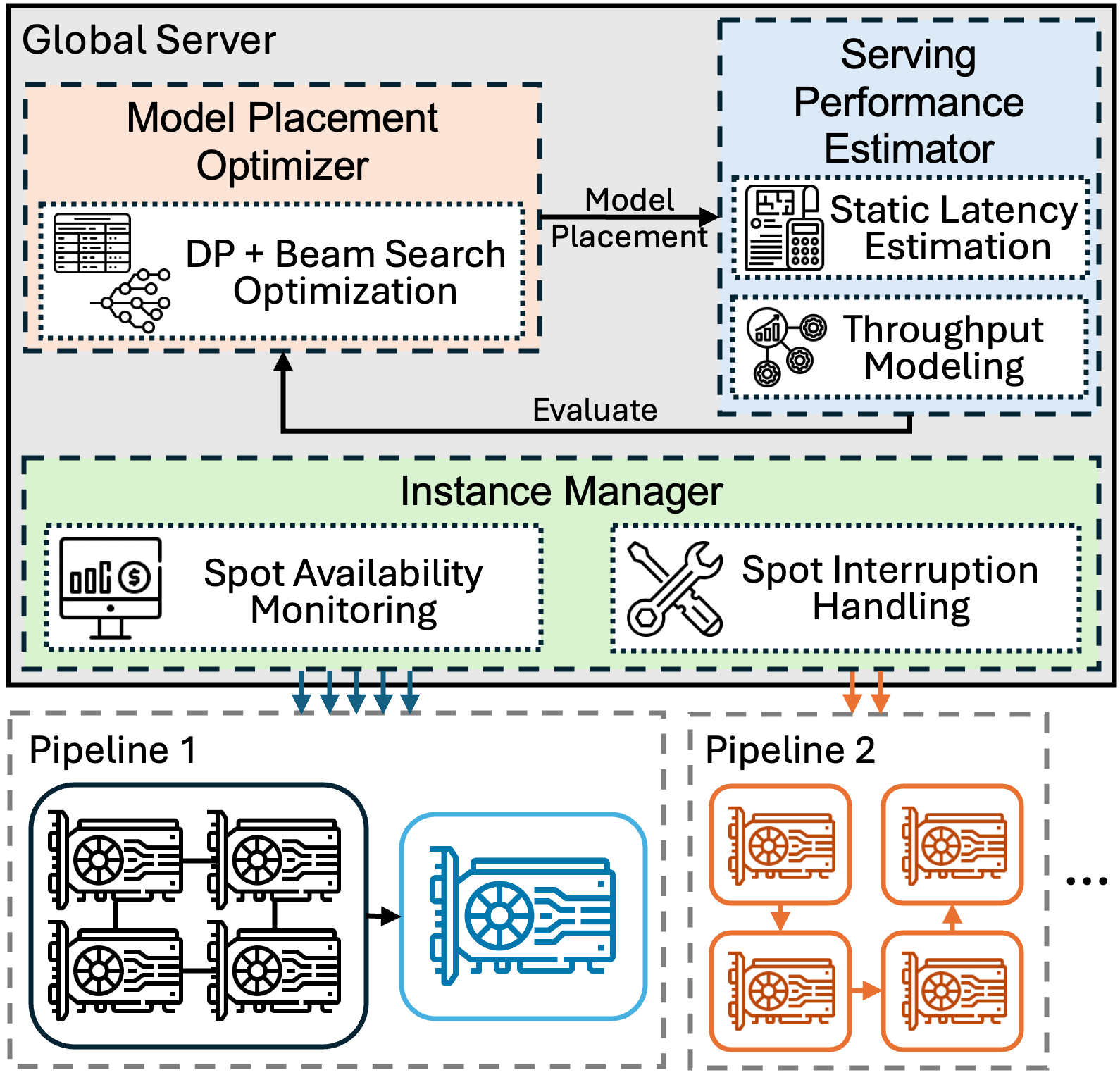}
\caption{ShuntServe System Overview.}
\label{fig:system_overview}
\end{figure}

\paragraph{The Optimization Challenge}
Although uneven layer partitioning and per-stage TP are effective partitioning strategies, determining the optimal model placement remains a complex challenge. LLM inference exhibits distinct performance characteristics where the prefill phase is typically compute-bound, and the decode phase is memory-bound~\cite{distserve,splitwise,thunderserve,pmlr-v267-jiang25c}. In a heterogeneous cluster, this behavior leads to decoupled bottlenecks, where the bottleneck stage differs between phases. As shown in Figure~\ref{fig:decoupled_bottleneck}, in a four-stage pipeline of heterogeneous GPUs, Stage 0 is the bottleneck during prefill, while Stage 1 becomes the bottleneck during decode. Maximizing throughput requires jointly optimizing a vast set of factors, such as GPU types, pipeline depth, TP degree, and layer allocation. The resulting search space can become exponentially large, rendering a naive exhaustive search intractable.

\section{ShuntServe Design Overview}
\label{sec:shuntserve_design}

Implementing cost-effective LLM serving on heterogeneous spot GPU clusters requires addressing two primary challenges. First, determining the optimal model placement in a heterogeneous GPU cluster is complicated by a vast search space of possible model placement configurations, rendering profiling and exhaustive search methods intractable. Second, a spot instance interruption triggers a system-wide pipeline stall, resulting in significant downtime and loss of progress for ongoing requests.

ShuntServe determines the optimal model placement on heterogeneous GPU clusters through a \emph{serving performance estimator} that derives the performance of diverse configurations, and a \emph{partitioned model placement optimizer}, which efficiently explores the combinatorially explosive search space of model placements (Section~\ref{sec:model_placement}). To address spot interruptions, ShuntServe combines \emph{output-preserving request migration} to recover interrupted in-flight requests with \emph{concurrent initialization} to minimize downtime by exploiting the grace period (Section~\ref{sec:fault_tolerance_design}).

Figure~\ref{fig:system_overview} shows an overview of ShuntServe. The global server serves as the master node of the cluster. It hosts three key components: the serving performance estimator, the model placement optimizer, and the instance manager. The serving performance estimator predicts the latency and throughput of a given model placement. The model placement optimizer searches for the optimal model placement through DP with beam search, retaining the top-$k$ model placements evaluated by the serving performance estimator in each DP entry. The instance manager monitors spot availability and invokes concurrent initialization and output-preserving request migration to handle interruption on the affected pipeline. Each pipeline consists of heterogeneous GPU instances, either spot or on-demand, and requests are dispatched via a basic weighted round-robin policy based on per-pipeline throughput.

\section{Model Placement for Heterogeneous GPUs}
\label{sec:model_placement}
This section addresses the method for finding the optimal model placement within the vast search space of a heterogeneous GPU cluster with the parallelization strategies. 

\subsection{LLM Serving Performance Estimator}
\label{subsec:serving_performance_estimator}
To make an optimal model layer placement on heterogeneous GPU devices, it is important to estimate the performance on disparate devices. Prior studies have relied on thorough empirical profiling to predict the performance of LLM inference systems \cite{helix,vidur,AlpaServe}. However, the factors influencing throughput, such as model parameter size, sequence length, batch size, and GPU type, create an extensive combinatorial space. This combinatorial explosion is further exacerbated when considering parallelization strategies of TP and PP, making the cost of exhaustive profiling intractable. 
Consequently, an approach that minimizes or eliminates profiling is essential for efficiently finding an optimal, cost-effective LLM serving configuration.

\begin{table*}[t]
\centering
\begin{threeparttable}
\caption{FLOPs and memory scan requirements for generative LLM inference.}
\label{tab:operations}
\begin{tabular}{llll}
\hline\hline
Operation                                                                        & Phase   & FLOPs & Memory Scan Cost  \\ 
\hline
\multirow{2}{*}{\begin{tabular}[c]{@{}l@{}}QKV\\Projection\end{tabular}}         & Prefill & $B(2S_{in}H^2+4S_{in}HH_{kv})/D_{TP}$ & $BS_{in}H+(H^2+2HH_{kv})/D_{TP}$  \\ 
                                                                                 \cline{2-4}
                                                                                 & Decode  & $BS_{out}(2H^2+4HH_{kv})/D_{TP}$ & $S_{out}(BH+(H^2+2HH_{kv})/D_{TP})$             \\ 
\hline
\multirow{2}{*}{Attention}                                                       & Prefill & $4BS_{in}^2H/D_{TP}$ & $(BS_{in}H+2BS_{in}H_{kv})/D_{TP}$  \\
                                                                                 \cline{2-4}
                                                                                 & Decode  & $\sum_{t=1}^{S_{out}} 4(s_{in}+t)H/D_{TP}$ & $\sum_{t=1}^{S_{out}} (BH+2B(S_{in}+t)H_{kv})/D_{TP}$           \\ 
\hline
\multirow{2}{*}{\begin{tabular}[c]{@{}l@{}}Output\\Projection\end{tabular}}      & Prefill & $2BS_{in}H^2/D_{TP}$ & $(BS_{in}H+H^2)/D_{TP}$  \\
                                                                                 \cline{2-4}
                                                                                 & Decode  & $2BS_{out}H^2/D_{TP}$ & $S_{out}(BH+H^2)/D_{TP}$  \\ 
\hline
\multirow{2}{*}{\begin{tabular}[c]{@{}l@{}}Up and Gate\\Projection\end{tabular}} & Prefill & $4BS_{in}HH_{ffn}/D_{TP}$ & $BS_{in}H+2HH_{ffn}/D_{TP}$  \\
                                                                                 \cline{2-4}
                                                                                 & Decode  & $4BS_{out}HH_{ffn}/D_{TP}$ & $S_{out}(BH+2HH_{ffn}/D_{TP})$  \\ 
\hline
\multirow{2}{*}{\begin{tabular}[c]{@{}l@{}}Down\\Projection\end{tabular}}        & Prefill & $2BS_{in}HH_{ffn}/D_{TP}$ & $(BS_{in}H_{ffn}+HH_{ffn})/D_{TP}$             \\
                                                                                 \cline{2-4}
                                                                                 & Decode  & $2BS_{out}HH_{ffn}/D_{TP}$ & $S_{out}(BH_{ffn}+HH_{ffn})/D_{TP}$  \\ 
\hline
\multirow{2}{*}{\begin{tabular}[c]{@{}l@{}}Logits\\Calculation\end{tabular}}     & Prefill & $2BS_{in}H\mathcal{V}/D_{TP}$ & $BS_{in}H+H\mathcal{V}/D_{TP}$             \\
                                                                                 \cline{2-4}
                                                                                 & Decode  & $2BS_{out}H\mathcal{V}/D_{TP}$ & $S_{out}(BH+H\mathcal{V}/D_{TP})$ \\
\hline\hline
\end{tabular}
\end{threeparttable}
\end{table*}

\begin{table}[h]
\centering
\caption{Notations}
\label{tab:notation}
\begin{tabular}{cl}
\toprule
Symbol & Description \\
\hline
$L$ & Latency \\
$B$ & Batch size \\
$S_{in}$ & Input sequence length \\
$S_{out}$ & Output sequence length \\
$H$ & Hidden dimension \\
$H_{kv}$ & Key/Value head dimension \\
$H_{ffn}$ & Intermediate dimension in FFN \\
$\mathcal{V}$ & Vocabulary size \\
$D_{PP}$ & Pipeline parallelism degree \\
$D_{TP}$ & Tensor parallelism degree \\
$l$ & Number of transformer layers \\
$E$ & Element size in bytes \\
$t$ & Iteration of current step \\
\bottomrule
\end{tabular}
\end{table}

\subsubsection{Static Latency Estimation}

To accurately calculate system throughput, one must first estimate inference latency. Following the roofline model~\cite{roofline, hierarchicalrooflineperformanceanalysis}, which dictates that an operation's latency is determined by the maximum of its compute latency and memory scan latency. Equation~\ref{eq:latency} formulates the latency of each operation based on its arithmetic intensity. The total end-to-end inference latency for a given device is then computed by summing the latencies of all operations it must perform. 

\begin{align}
    &L_{compute} = \frac{FLOPs}{FLOPS} \nonumber \\
    &L_{memory} = \frac{{Mem Scan Cost}\cdot E}{\textit{Mem. BW.}} \nonumber \\
    &L_{ops} = \max(L_{compute},L_{memory})
    \label{eq:latency}
\end{align}

The operations are formulated based on the three components in which most computations occur: (1) Attention, (2) Feed Forward Network (FFN), and (3) Output Embedding. These components are the de-facto standard of transformer-based generative LLMs, and the formulations therefore generalize to other models that adopt the same structure. Table~\ref{tab:operations} details the FLOPs and memory scan cost required for each defined operation. This formulation-based static latency estimation derives latency for any pipeline configuration through fast calculation, applying flexibly to new configurations without requiring prior profiling. Table~\ref{tab:notation} summarizes the notation used throughout this paper.

\subsubsection{Communication Overhead Modeling}
To estimate the end-to-end inference latency in a distributed serving system, network latency should be considered alongside the compute latency. ShuntServe models the network latency for pipeline communication using the $\alpha$-$\beta$ model \cite{HOCKNEY1994389, improving_the_performance_of_collective_operations_in_mpich}, which estimates communication cost as the sum of a constant network latency $\alpha$ and the data transfer time proportional to message size over bandwidth $\beta$, and the communication topology for TP operations, e.g., AllReduce/AllGather, is modeled using a Ring-based \cite{hu2025nccl} approach. These communication latency costs are factored into the calculation for each stage and phase.

In the Ring-based topology, a single \textit{AllReduce} operation consists of one \textit{ReduceScatter} and one \textit{AllGather} operation, where each operation transmits data of size $N/P$ for $P-1$ iterations. This can be expressed as $(\alpha+\frac{N/P}{\beta})(P-1)$. Here, $\alpha$ and $\beta$ denote network latency and bandwidth, respectively, $N$ represents the size of the original data to be shared, and $P$ is the number of processes participating in the communication, which corresponds to $D_{TP}$ in TP communication. These AllReduce operations occur twice per transformer layer, and we ultimately calculate the PP and TP communication latencies as follows:

\begin{align}
    &L_{PP}=\alpha_\text{inter}+\frac{BSHE}{\beta_\text{inter}} \\
    &L_{TP}=4\cdot(\alpha_\text{intra}+\frac{BSHE}{D_{TP}\cdot \beta_\text{intra}})\cdot(D_{TP}-1)\cdot l
\end{align}

$\alpha_\text{inter}$ and $\beta_\text{inter}$ denote the inter-stage (e.g., Ethernet, InfiniBand) network latency and bandwidth, and $\alpha_\text{intra}$ and $\beta_\text{intra}$ denote the intra-stage (e.g., PCIe, NVLink) network latency and bandwidth.

\subsubsection{Throughput Modeling}
While the latency of a single request is critical, the total number of requests the system can process per unit of time, throughput, is an equally important performance metric for a production LLM serving system. This is often measured in Requests per Second (RPS) and can be expressed as Equation \ref{eq:rps}.

\begin{equation}
    \text{RPS} = \frac{B}{\sum{L_{ops}(B)}}
    \label{eq:rps}
\end{equation}

ShuntServe models the throughput of a heterogeneous pipeline by first using the static latency estimator to calculate the distinct prefill and decode latencies for each stage $i$. The overall throughput is then derived by computing the latency using Equation~\ref{eq:throughput_modeling} and then combining it with Equation~\ref{eq:rps}.

\begin{equation}
\begin{split}
    % L(B) = \max_{0\le i<D_{PP}}(L^i(B)) + \max_{0\le i<D_{PP}}(\sum_{t=1}^{S_{out}}L^i_t(B))
    L(B) = \max_{0\le i<D_{PP}}(L^i_\text{prefill}(B)) + \max_{0\le i<D_{PP}}(L^i_\text{decode}(B))
\end{split}
\label{eq:throughput_modeling}
\end{equation}

Such formulation-based performance models require hardware characteristics such as FLOPS and memory bandwidth. We initially relied on the official white papers of each GPU and instance, but empirically found that the reported values diverge substantially from the effective performance observed on actual hardware~\cite{habitat, flash_attention2}. ShuntServe therefore performs a one-time calibration through operations that saturate distinct hardware characteristics (i.e., compute-bound GEMM, memory-bound GEMV, and network-bound AllReduce), and we evaluate its overhead in Section~\ref{eval:calibration_overhead}.

\subsection{Partitioned Model Placement Optimizer} \label{subsec:model_placement_optimizer}

ShuntServe determines the optimal model placement using the throughput modeling described in the previous section. The most intuitive approach, an exhaustive search over all possible placements, is computationally intractable as the search space is exponentially large. To address this, ShuntServe employs a DP algorithm augmented with beam search that efficiently navigates this search space. In this section, we describe the details of the algorithm, the rationale for its batch size determination, and the optimization objective.

\subsubsection{Model Placement with DP} \label{subsubsec:model_placement_algorithm_dynamic_programming}

A key capability of ShuntServe's approach is that the operation-level latency estimation and throughput modeling from Section~\ref{subsec:serving_performance_estimator} enable evaluating partial model placements within DP subproblems, exploiting the optimal substructure property. Following prior studies~\cite{thunderserve,helix,hexgen} and our empirical experience, TP is restricted to intra-node GPUs, as inter-node network latency and bandwidth is insufficient. Furthermore, in the spot environment, ShuntServe assigns each instance exclusively to a single pipeline, as sharing an instance across multiple pipelines might cause failures where the impact of interruption propagates to all pipelines containing it.

\begin{algorithm}[tb]
    \caption{Partitioned Model Placement Optimizer}
    \label{alg:model_placement_optimizer}
\begin{algorithmic}[1]
    \REQUIRE Beam size $k$ (default 1), instance types $\mathcal{I}$, total layers $N_L$
    \ENSURE Model Placement $\mathcal{M}$
    \STATE Initialize $DP[l][s] \leftarrow \emptyset$ for all $l \in [0, N_L], s \in [0, N_L]$
    \FOR{$l = 1$ {\bfseries to} $N_L$}
        \FOR{$l' = 0$ {\bfseries to} $l-1$}
            \STATE $l_{\text{new}} \leftarrow l - l'$
            \FOR{$s = 0$ {\bfseries to} $l'$}
                \STATE $s_{\text{new}} \leftarrow s + 1$
                \STATE $\mathcal{C} \leftarrow$ Top-$k$ stage combinations from $DP[l'][s]$
                \FOR{$(c, i) \in \mathcal{C} \times \mathcal{I}$}
                    \STATE $c_{\text{new}} \leftarrow c.\texttt{add\_stage}(i, l_{\text{new}})$
                    \STATE Compute maximum batch size $B_{\text{max}}$ for $c_{\text{new}}$
                    \STATE Evaluate throughput of $c_{\text{new}}$
                    \STATE Update $DP[l][s_{\text{new}}]$ with $c_{\text{new}}$ if better
                \ENDFOR
            \ENDFOR
        \ENDFOR
    \ENDFOR
    \STATE $\mathcal{M} \leftarrow \arg\max_{c \in DP[N_L][*]} \text{objective}(c)$
\end{algorithmic}
\end{algorithm}

Algorithm~\ref{alg:model_placement_optimizer} formalizes this optimization approach. The problem of placing $l$ layers across $s+1$ stages is decomposed into a subproblem of placing $l'<l$ layers across $s$ stages and the assignment of the remaining $l-l'$ layers to a new stage with a specific instance type (lines 2--4). For each prior configuration in $DP[l'][s]$, the configuration is extended with each available instance type (lines 8--9). The maximum batch size is then computed for the resulting placement under memory constraints (line 10), and the throughput is estimated using the serving performance estimator (line 11). The DP table $DP[l][s+1]$ is updated if the new configuration achieves a higher score (line 12), and after all layers are placed, the optimal placement is extracted from $DP[N_L][*]$ (line 17).

Even with a DP approach, the search space for model placement can grow exponentially with the number of configurable stages and instance types. The number of stages can, in theory, be as high as the total number of layers, $N_L$, and each stage can be assigned one of $|\mathcal{I}|$ instance types. When including the combinations for layer allocation, the time complexity of exhaustive DP implementation reaches $O(N_L^3 \cdot N_S!/\Pi_{i\in \mathcal{I}}\binom{N_S}{|i|})$, which becomes computationally infeasible. To manage this, ShuntServe incorporates beam search, inspired by top-$k$ token selection in autoregressive generation~\cite{graves2012sequencetransductionrecurrentneural, seq-to-seq}, retaining only the top-$k$ partial placement candidates in each $DP[l][s]$ entry (line 7). This optimization prunes the search space, dramatically reducing the time complexity to $O(N_L^3 \cdot k \cdot |\mathcal{I}|)$. ShuntServe employs this efficient optimization process iteratively, allowing it to greedily extract the desired number of pipeline configurations to populate the serving system.

\subsubsection{Batch Size Determination}\label{subsubsec:batch_size_determination}
In LLM serving, increasing the batch size significantly improves throughput. This efficiency stems from a disparity between the inference phases. Latency in the compute-bound prefill phase scales linearly with batch size, whereas latency in the memory-bound decode phase exhibits minimal growth, as multiple requests in a batch share the same weight matrix and require only a single memory access, making batching highly efficient in the decode phase. Since the decode phase typically dominates the overall request latency~\cite{splitwise,vllm,sanovar2025leanattention,characterizing_power_management_opportunities_patel}, this batching efficiency directly translates into throughput improvements. Thus, ShuntServe maximizes serving throughput by setting the largest possible batch size $B$ that satisfies the memory constraint of a given model placement, as in Equation~\ref{eq:memory_constraint}. 

\begin{equation}
    B=\min_{0\le i<D_{PP}}(\lfloor\frac{M^i-M^i_{\text{weight}}-M_{\text{activation}}}{l^i\cdot(S_{in}+S_{out})}\rfloor)
    \label{eq:memory_constraint}
\end{equation}

In the equation, $M^i$, $M^i_{\text{weight}}$, and $M_{\text{activation}}$ denote the total memory capacity of stage $i$, the model weight size, and the maximum activation tensor size during a forward pass, respectively. Using the determined batch size $B$, the model placement optimizer evaluates the maximum throughput of the given placement.

\subsubsection{Optimization Objective}\label{subsubsec:optimization_objective}
Using throughput as the sole optimization objective is problematic. Adding instances increases the available memory, which in turn allows for larger batch sizes and higher throughput. This creates an incentive for the algorithm to indiscriminately add instances, regardless of cost-efficiency. To prevent this, ShuntServe defines its objective function as throughput per cost, thereby evaluating the cost-effectiveness of each potential instance addition. We formulate this as:

\begin{equation}
    \underset{\mathbf{p}}{\arg\max}(\frac{\text{Throughput}(\textbf{p})}{\text{Cost}(\textbf{p})}\cdot(1-\gamma\cdot \max(0, \frac{\text{latency}(\textbf{p})}{SLO}-1)))
\end{equation}

$\mathbf{p}$ denotes the model placement of a pipeline. We primarily optimize for throughput per cost, but include a soft penalty term for latency to satisfy the desired Service Level Objective (SLO). The parameter $\gamma$ controls the sensitivity to this latency penalty; setting $\gamma = \infty$ enforces it as a hard constraint. In the ShuntServe default setup, $\gamma$ is set to zero to prioritize throughput maximization. This removes the explicit latency constraint but does not imply a sacrifice in latency, as throughput and latency are closely coupled. We evaluate the resulting latency behavior under this setting in Section~\ref{eval:online_model_placement}.

\section{Handling Spot Interruption}
\label{sec:fault_tolerance_design}
Effectively solving fault tolerance challenge raised from spot instance interruption requires addressing two specific sub-problems. First is how to preserve the state of ongoing inference requests. In autoregressive decoding, simply discarding the generated tokens and KV cache wastes all prior computation and results in the loss of the user's generated output. Second is how to rapidly reconfigure the pipeline with new resources substituting interrupted ones. ShuntServe addresses these sub-problems through \emph{output-preserving request migration} and \emph{concurrent initialization} based on a shared tensor store.

\begin{figure}[t]
    \centering
    \includegraphics[width=0.75\linewidth]{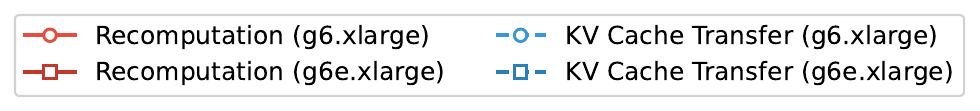}
    \vspace{-0.5em}
    
    \raisebox{0.5cm}{\includegraphics[height=2cm]{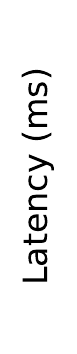}}%
    \hspace{-0.25em}%
    \subfloat[3B.\label{fig:migration_3b}]{%
        \includegraphics[width=0.32\linewidth]{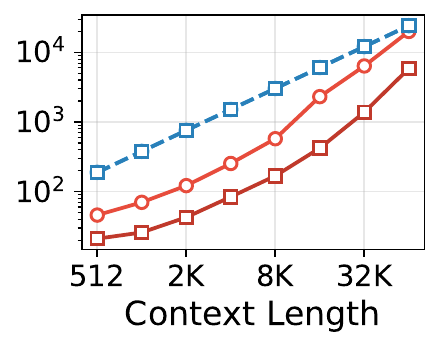}}%
    \subfloat[8B.\label{fig:migration_8b}]{%
        \includegraphics[width=0.32\linewidth]{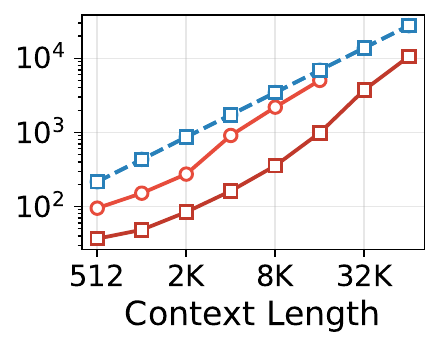}}%
    \subfloat[70B (per-layer).\label{fig:migration_70b}]{%
        \includegraphics[width=0.32\linewidth]{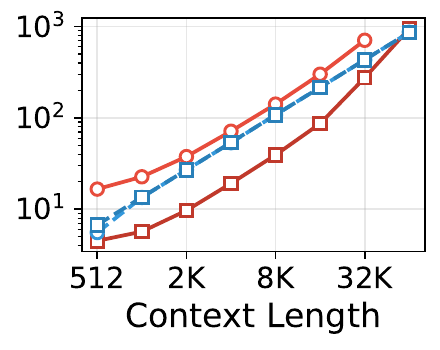}}
    
    \caption{KV cache transfer vs recomputation latency on Llama 3 family (3B, 8B, 70B).}
    \label{fig:migration_comparison}
\end{figure}

\subsection{Output-Preserving Request Migration}
\label{subsubsec:request_migration}
Rather than simply discarding the generated tokens and KV cache, two primary approaches exist to migrate requests while preserving their computation progress. The first is KV cache transfer-based migration, which transfers the KV cache from the interrupted instances to new replacement instances, allowing the receiving instances to resume inference requests without recomputation. The second is recomputation-based migration, which routes the interrupted requests' input sequence and already-generated output sequence to a pipeline containing new instances to reconstruct the KV cache.

Figure~\ref{fig:migration_comparison} shows the latency overhead of the two approaches measured on AWS g6.xlarge (L4) and g6e.xlarge (L40S) instances across varying context lengths for the Llama~3 family with 3B, 8B and 70B model sizes. The x-axis represents the context length, and the y-axis represents latency on a log scale. For the 70B model, since all layers cannot fit on a single instance, we measured the latency of a subset of layers and normalized it to a per-layer basis. The results show that recomputation-based migration exhibits lower latency overhead in most cases, demonstrating the advantage of the recomputation-based approach.

On the other hand, KV cache transfer can be effective in certain environments (i.e., very large models, very long contexts, GPUs with limited compute capacity, and high-bandwidth networks). For example, in the case of Llama-3-70B (Figure~\ref{fig:migration_70b}), recomputation exhibited higher latency than transfer across all context lengths on L4. On L40S, recomputation showed lower latency up to 32K context length, but at 64K it incurred a recomputation latency of 7.59s compared to 6.92s for transfer, resulting in approximately 9.6\% higher latency. 

However, in spot instance environments, KV cache transfer introduces challenges beyond latency, particularly in terms of fault tolerance. If an interruption occurs during KV cache transfer, the system must ultimately fall back to recomputation, incurring the overhead of both approaches. Moreover, spot interruption environments impose a strict constraint that the entire process, from new instance initialization to transfer completion, must be completed within the grace period, rendering the transfer-based approach impractical for spot environments. Therefore, ShuntServe adopts the recomputation-based approach, which is practical under spot instance constraints. 

\begin{figure}[t]
\centering
\includegraphics[width=0.45\textwidth]{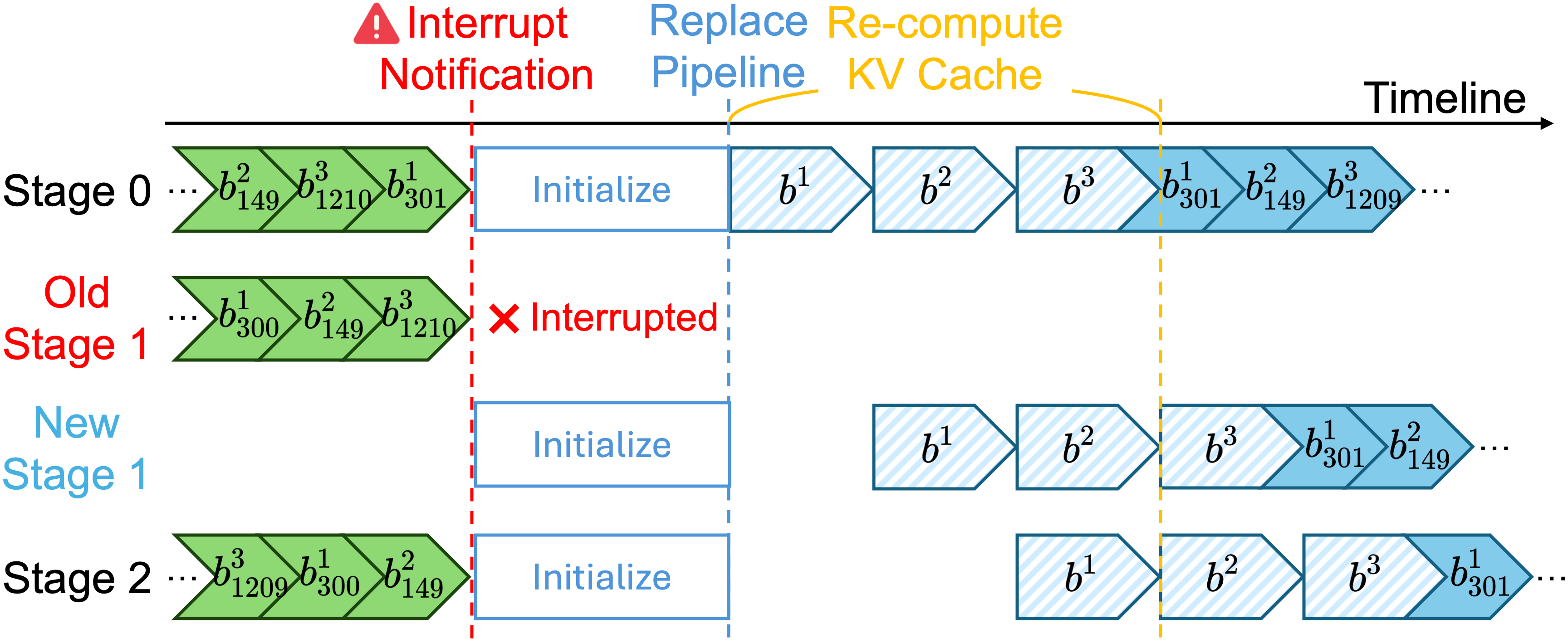}
\caption{Output-preserving request migration. $b^j_t$ denote the $t$-th iteration of the $j$-th micro-batch.}
\label{fig:output_preserving_request_migration}
\end{figure}

Figure~\ref{fig:output_preserving_request_migration} illustrates an output-preserving request migration scenario in which Stage 1, among three stages, is interrupted. Once the new pipeline is initialized, in-flight requests are recomputed while preserving their original inputs and previously generated outputs (e.g., in the figure, the three micro-batches are interrupted at iterations 301, 149, and 1209, respectively). The interrupted micro-batches subsequently recover their KV caches through recomputation, the cost of which is substantially lower than the cumulative iteration time required to generate the previously emitted output tokens, thereby enabling in-flight requests to be efficiently migrated to the new pipeline.

% =======================

\subsection{Minimizing Downtime via Concurrent Initialization}
\label{subsubsec:concurrent_initialization}
Selectively replacing only the interrupted node while continuing inference on the unaffected nodes reduces resource waste and downtime. However, this partial pipeline replacement introduces technical challenges unsolved by existing LLM serving systems. When replacing only part of a pipeline, the replacement node must be added to the cluster and the distributed inference engine must be re-initialized with the unaffected nodes.

The core problem is that existing LLM serving systems, such as vLLM \cite{vllm} and TensorRT-LLM \cite{nvidia2023tensorrtllm}, tightly couple the inference engine lifecycle with model weight management. Ideally, the new pipeline incorporating the replacement node would be initialized while the existing pipeline continues serving requests, requiring both old and new engine processes to temporarily coexist on each unaffected node. However, because each engine process independently loads its own copy of the model weights and allocates KV cache space in GPU memory, this coexistence results in duplicate memory allocation that triggers an out-of-memory (OOM) error. Consequently, the new pipeline can only start after the old one is fully terminated, precluding any overlap between preparation and serving and incurring significant downtime.

\paragraph{Shared Tensor Store} To resolve inefficiencies arising from the coupled lifecycle, ShuntServe manages model weights and KV caches in a separate \emph{Shared Tensor Store} process, decoupling the lifecycle of the inference engine from that of the model. In the Shared Tensor Store, model weight tensors and KV cache space are shared with the inference engine process in a zero-copy manner through CUDA IPC handles, allowing the engine to directly reference the tensors during inference. This eliminates copy and communication overhead during inference, and prevents duplicate memory allocation upon re-initialization, as a new engine process can be attached to the existing store.

\begin{figure}[t]
\centering
\includegraphics[width=0.45\textwidth]{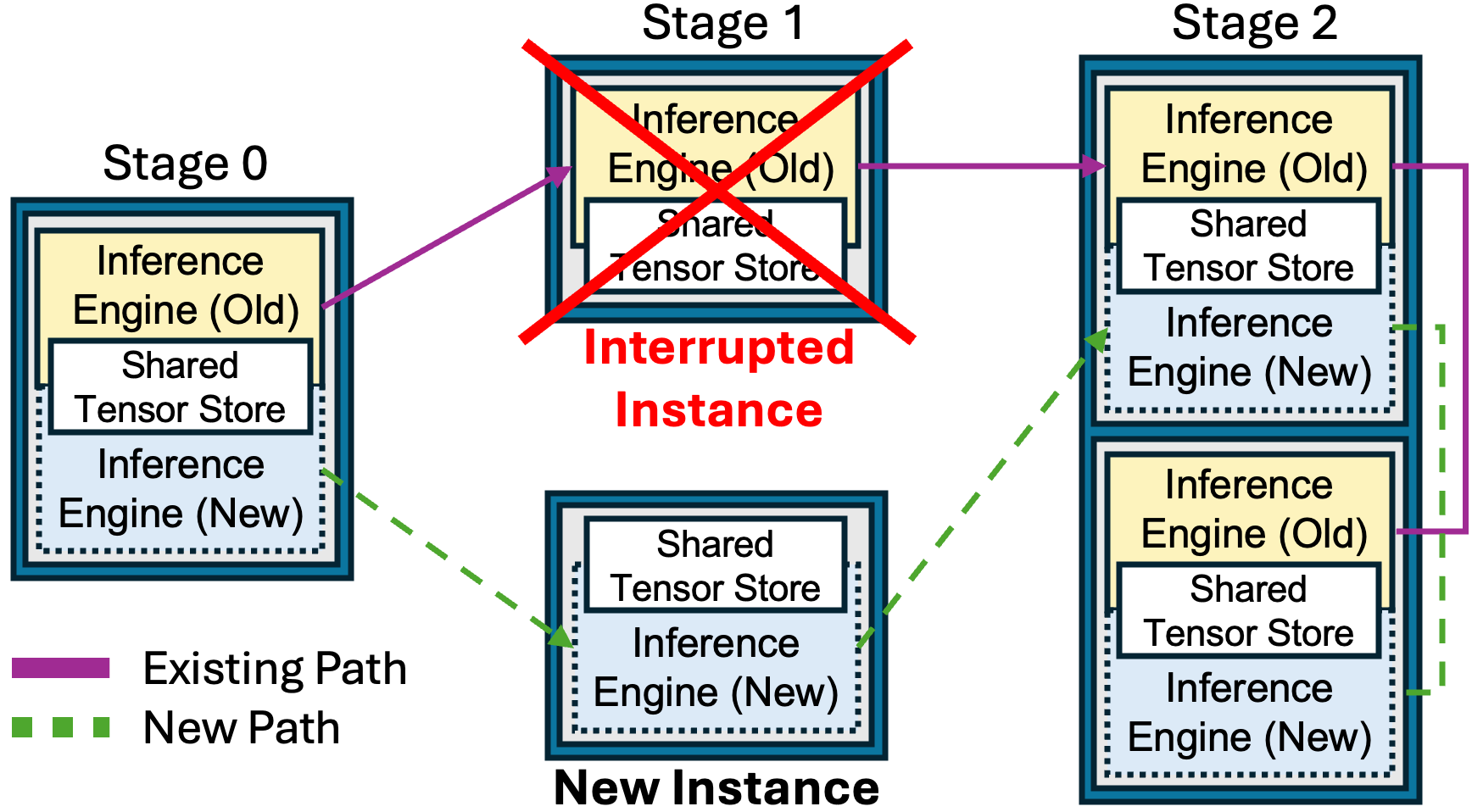}
\caption{Concurrent initialization with shared tensor store. }
\label{fig:concurrent_initialization}
\end{figure}

Figure~\ref{fig:concurrent_initialization} illustrates the concurrent initialization in ShuntServe. When the global server detects a spot interruption, it adds the new instance to the cluster and starts a \emph{Shared Tensor Store} process on the new instance. It simultaneously loads the model partition from remote storage and initializes inference engines across the new instance and the unaffected instances. This preparation occurs in the background while the existing pipeline continues serving.

Once the initialization of the new pipeline is complete, the global server stops routing requests to the old pipeline, terminates its inference engine, and shunts the request path to the new pipeline. Since the new pipeline initialization overlaps with ongoing serving, the transition is achieved with near-zero downtime when the initialization completes within the grace period. Note that even when the initialization time exceeds the grace period, downtime is incurred only for the portion that extends beyond the grace period.

\section{Implementation}
\label{sec:implementation}
The ShuntServe inference engine was implemented by modifying vLLM \cite{vllm} v0.8.1. The modifications comprise 1,077 lines of Python code. These modifications include changes to the communication methods to support asymmetric parallelism and code for fetching model weights from the Shared Tensor Store. The Shared Tensor Store itself, along with the global server and scheduler for managing pipelines, were written in 10,363 lines of Python code. The scheduler implements load balancing via a weighted round-robin policy based on the throughput of each pipeline.

The global server uses Ray \cite{ray} to manage physical instances, handling their participation in the cluster and subsequent eviction. Model loading is handled by downloading from a remote object storage server. To minimize loading time, model tensors are sharded into individual binary files in a custom raw binary format. In our observation, \texttt{torch.save()} retains the full original tensor size when saving sliced tensors, resulting in significant network overhead for downloading partitioned shards. The custom binary format eliminates this overhead by storing only the actual partition data. This design allows each distributed node to download only its required partition via multi-threading and load it directly onto the GPU.

\section{Evaluation}
\label{sec:evaluation}
This section comprehensively evaluates the key contributions of ShuntServe. The evaluation is structured into two main parts: an assessment of the model placement optimizer's performance on heterogeneous GPU clusters (Section~\ref{eval:model_placement}), and an evaluation of the fault tolerance mechanism's effectiveness under spot interruptions (Section~\ref{eval:interruption_tolerance}).

\paragraph{Model and Cluster Setup} ShuntServe is evaluated using the half-precision (BF16) Llama-3.1-70B-Instruct~\cite{grattafiori2024llama3herdmodels} and Qwen3-32B~\cite{qwen3} models. The evaluation cluster consists of heterogeneous AWS GPU instances: three g6.12xlarge instances (each with 4x L4 24GB GPUs), two g5.12xlarge instances (each with 4x A10G 24GB GPUs), and four g6e.xlarge instances (each with 1x L40S 48GB GPU). This provides a total of 24 GPUs and 672GB of GPU memory. The evaluation experiments were conducted on AWS, consuming approximately 1,000 GPU-hours in total, excluding costs for development and debugging.

\paragraph{Request Scenario} We use the Azure Conversation Dataset~\cite{splitwise}, a production trace from Azure LLM inference services, which spans one hour with fluctuating request arrival patterns and input/output length distributions. We prune requests with input lengths exceeding 2048 tokens, as peak context lengths in the dataset cause OOM errors on homogeneous baselines whose memory of each pipeline is limited, rendering comparison infeasible. The resulting dataset has an average input length of 763 and an average output length of 232, and an average request rate of 4.67 req/s. As also noted in~\cite{helix}, this distribution is more challenging than those used in prior work~\cite{vllm} due to its longer request lengths.

\begin{figure*}[t]
  \centering
  \includegraphics[width=\textwidth]{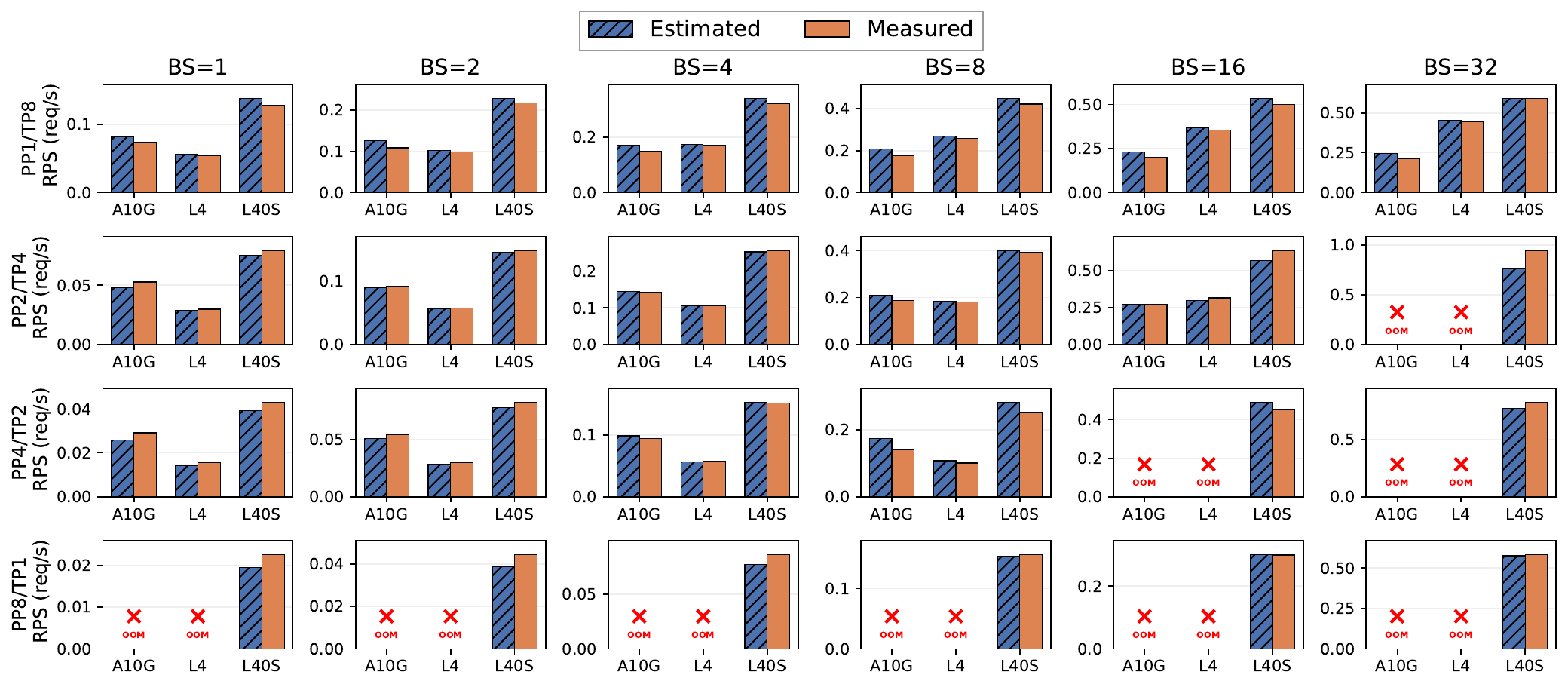}
  \caption{Comparison of estimated and measured inference latency for Llama-3.1-70B-Instruct across parallelism configurations (PP$\times$TP), GPU types (A10G, L4, L40S), and batch sizes (BS). Each row corresponds to a parallelism strategy, each column to a batch size, and the x-axis within each subplot represents the GPU type. Red crosses indicate OOM cases.}
  \label{fig:estimator_accuracy}
\end{figure*}

\paragraph{Metrics} We evaluate serving performance under two workload types. In offline workload, where the system is fully saturated with requests, we measure throughput in terms of RPS. In online workload, where requests arrive dynamically and the system processes them in real time, we measure request latency using Time to First Token (TTFT), which captures the duration from request dispatch through prefill to the generation of the first output token, Time per Output Token (TPOT), which measures the average generation time of each token during the decode phase, and end-to-end latency, which captures the total duration from request dispatch to completion. In the fault tolerance evaluation, we additionally compare the monetary cost for operating each system. We define cost efficiency as cost per throughput for offline workloads and as request latency$\times$cost for online workloads.

\subsection{Model Placement Evaluation}\label{eval:model_placement}

We evaluate the model placement optimizer of ShuntServe, which determines the optimal node assignment and layer partitioning for inference pipelines across a heterogeneous GPU cluster. We first assess the accuracy of the static latency estimation technique, a foundational component of the optimizer (Section~\ref{eval:estimation_accuracy}). We then evaluate the performance of the model placement decisions produced by ShuntServe under offline (Section~\ref{eval:offline_model_placement}) and online (Section~\ref{eval:online_model_placement}) workloads. Finally, we conduct a sensitivity analysis on the beam search width $k$ with respect to optimizer execution time and the quality of the resulting placement (Section~\ref{eval:top_k_sensitivity_analysis}).

\subsubsection{Serving Performance Estimation Accuracy}\label{eval:estimation_accuracy}

To validate the accuracy of ShuntServe's roofline model-based performance estimator, we compare predicted inference latency against measured values on three GPU instance types (g5.48xlarge with 8 A10G, g6.48xlarge with 8 L4, and g6e.48xlarge with 8 L40S) across four parallelism strategies (PP1/TP8, PP2/TP4, PP4/TP2, PP8/TP1) and six batch sizes (1, 2, 4, 8, 16, 32). The input and output sequence lengths are set to 763 and 232, respectively, matching the average of the evaluation dataset. To isolate GPU execution performance from CPU scheduling overhead, we use the \texttt{gptBench} benchmark from TensorRT-LLM~\cite{nvidia2023tensorrtllm}, which employs static batching with automatically optimized kernels. We adopt Mean Absolute Percentage Error (MAPE) as the accuracy metric across all feasible configurations, excluding cases where the configuration results in OOM errors.

Figure~\ref{fig:estimator_accuracy} presents the comparison across all GPU types, parallelism strategies, and batch sizes. The estimator achieves an average MAPE of 3.49\% on L4, 6.68\% on L40S, and 9.67\% on A10G, yielding an overall MAPE of 6.63\% across 54 feasible configurations. The minimum error is 0.03\% (PP1/TP8, batch size 32, L40S) and the maximum is 23.72\% (PP2/TP4, batch size 32, L40S). This variance arises because the static latency estimation assumes linear scaling of each operation, whereas actual GPU execution involves non-linearity that the model does not fully capture. We further discuss this limitation and potential refinements in the Discussion (Section~\ref{sec:discussion}).

Despite this variance, the achieved overall MAPE of 6.63\% demonstrates sufficient accuracy as the foundational component of the model placement optimizer, while requiring only a one-time hardware calibration (FLOPS, memory bandwidth, and network bandwidth) per GPU type rather than per-configuration profiling.

\subsubsection{Offline Serving Throughput across Placement Algorithms}\label{eval:offline_model_placement}

% Offline: 1행 2열 (single column figure)
\begin{figure}[t]
\centering
\includegraphics[width=0.7\columnwidth]{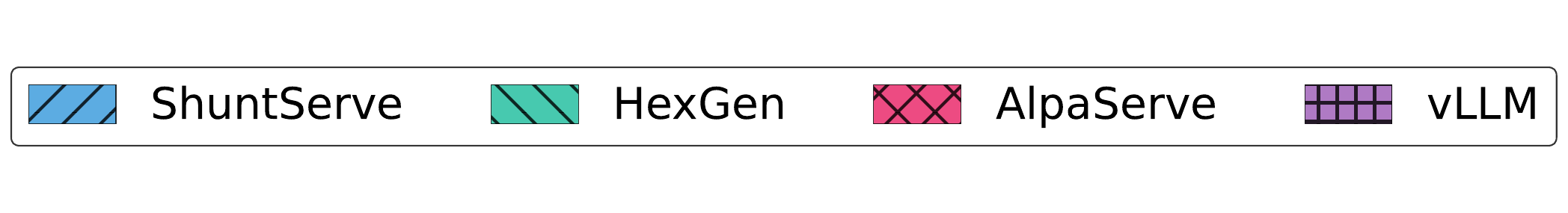}\\[-3mm]
\subfloat[Llama-3.1-70B\label{fig:model_placement_offline_llama}]{%
    \includegraphics[width=0.48\columnwidth]{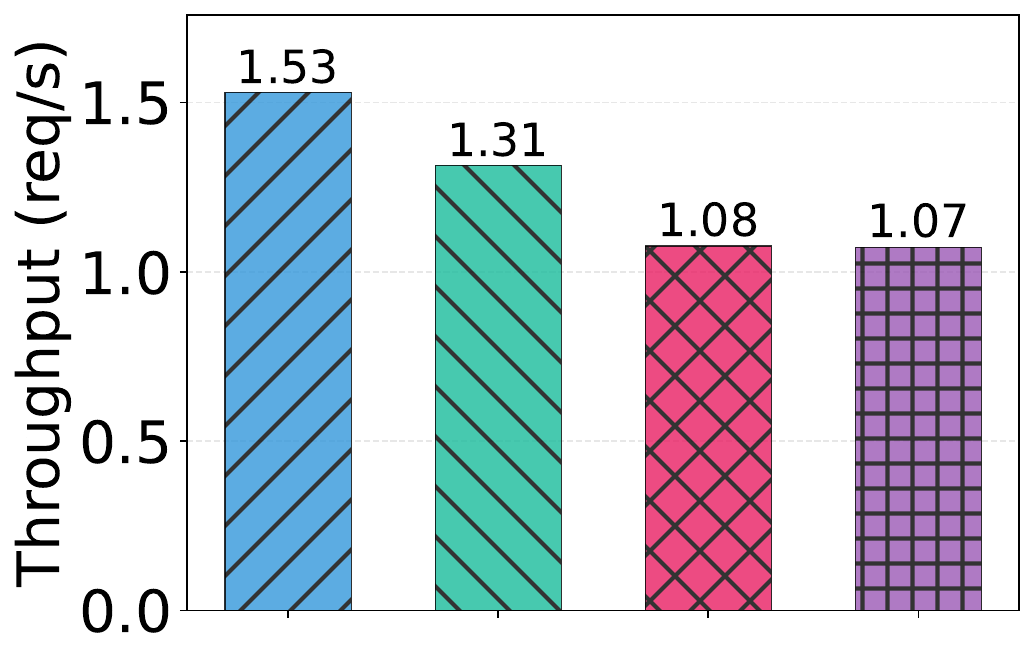}}
\hfill
\subfloat[Qwen3-32B\label{fig:model_placement_offline_qwen}]{%
    \includegraphics[width=0.48\columnwidth]{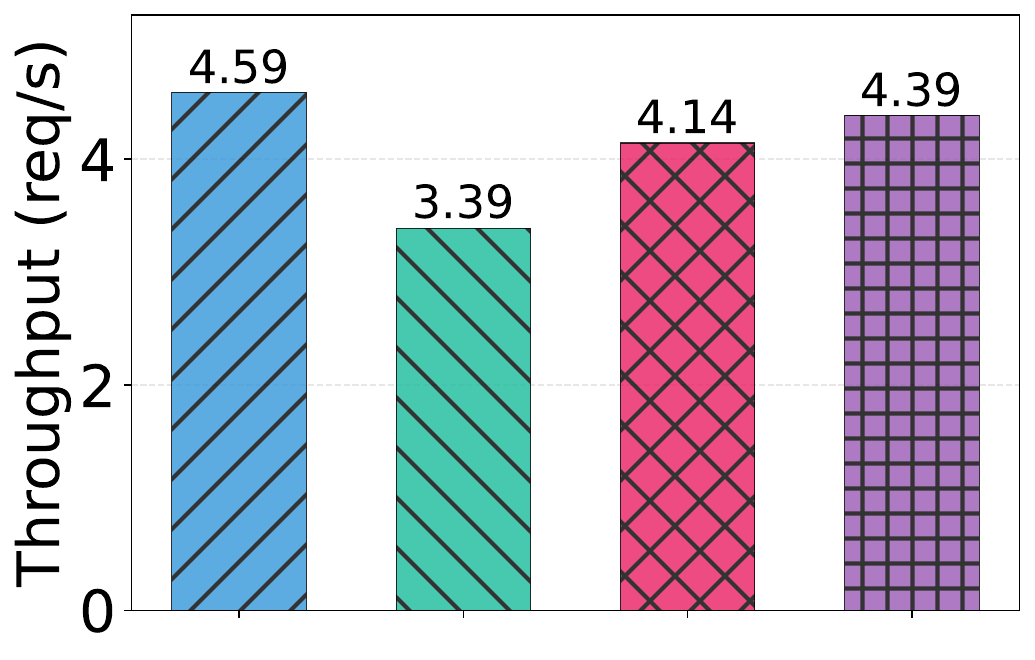}}
\caption{Offline throughput comparison under pipeline configurations determined by each baseline system's model placement algorithm. (a) shows results for Llama-3.1-70B, and (b) shows results for Qwen3-32B.}
\label{fig:model_placement_offline_throughput}
\end{figure}

We evaluate the model placement optimizer of ShuntServe by comparing against three basline systems. The first is HEXGEN~\cite{hexgen}, a state-of-the-art system for heterogeneous GPU clusters that uses a genetic algorithm and DP for model placement. The second is AlpaServe~\cite{AlpaServe}, a state-of-the-art system for homogeneous GPU clusters that uses DP to equalize stage latencies. The third, \emph{vLLM}~\cite{vllm}, is a widely adopted serving framework that applies even layer partitioning across homogeneous GPU clusters.

To measure the throughput under saturated conditions while prohibitive experimental cost, we use the first 5 minutes of the trace for Llama-3.1-70B (1,309 requests) and the first 20 minutes for Qwen3-32B (5,299 requests), as the smaller model requires a longer trace to maintain saturation. Figure~\ref{fig:model_placement_offline_throughput} presents the results. 

For Llama-3.1-70B (Figure~\ref{fig:model_placement_offline_llama}), ShuntServe achieves the highest throughput of 1.53 req/s, outperforming all baselines. In particular, ShuntServe improves throughput by 1.17$\times$ over HexGen, the state-of-the-art heterogeneous baseline. This improvement is attributed to the DP-based optimizer of ShuntServe, which explores the layer distribution space using roofline-based throughput estimation that accounts for the distinct characteristics of prefill and decode phases. In contrast, HexGen explores layer distributions through local perturbation within its genetic algorithm, which can become trapped in local optima when GPU capabilities are not ordered along the pipeline. Although HexGen mitigates this by distributing layers proportionally to the memory capacity of each stage, memory capacity does not necessarily correlate with GPU compute performance, leaving the resulting placement suboptimal.

Nevertheless, in Llama-3.1-70B, HexGen and ShuntServe, both heterogeneous systems, outperform the homogeneous baselines (AlpaServe and vLLM). This advantage stems from the ability of heterogeneous systems to flexibly compose pipelines from diverse GPU types, allowing each pipeline to retain sufficient memory for KV cache. For example, ShuntServe and HexGen each construct two pipelines for the cluster. In contrast, AlpaServe and vLLM construct three pipelines by grouping each GPU type separately, leaving insufficient memory for KV cache in each pipeline.

For Qwen3-32B (Figure~\ref{fig:model_placement_offline_qwen}), the results show a different trend. ShuntServe still achieves the highest throughput at 4.59 req/s, but the improvement over vLLM, which applies the intuitive even-partitioning placement in a homogeneous configuration, is only 1.05$\times$, compared to 1.43$\times$ for Llama-3.1-70B. This narrow gap arises from the evaluation cluster providing sufficient memory capacity for each pipeline even under homogeneous configurations for the smaller Qwen3-32B. In contrast, HexGen initializes pipeline groups based on communication topology, inherently producing heterogeneous configurations that persist through subsequent mutations. As a result, HexGen yields the lowest throughput among all baselines at 3.39 req/s. AlpaServe achieves 4.14 req/s, slightly below vLLM, as its SLO-oriented design selects additional replication. This increases the number of pipelines but reduces the batch size of each pipeline.

This contrast with the Llama-3.1-70B results indicates that when the cluster provides sufficient resources relative to the model size, an intuitive homogeneous configuration can be effective. ShuntServe's optimizer reflects this adaptability. For Qwen3-32B, it constructs each pipeline homogeneously to minimize decoupled bottlenecks, and differs from vLLM only in selecting replication over PP for the g5.12xlarge instance group based on throughput per cost.

\subsubsection{Analysis of Online Serving Latency}\label{eval:online_model_placement}

\begin{figure}[t]
\centering
\includegraphics[width=0.9\columnwidth]{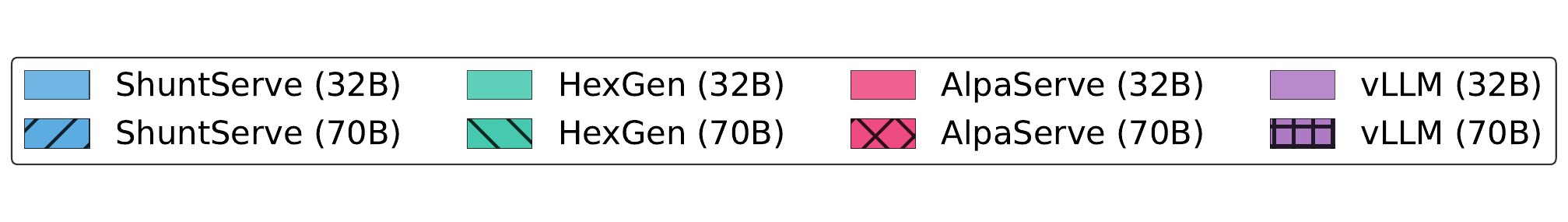}\\[-3mm]
\subfloat[TTFT\label{fig:model_placement_online_ttft}]{%
    \includegraphics[width=0.48\columnwidth]{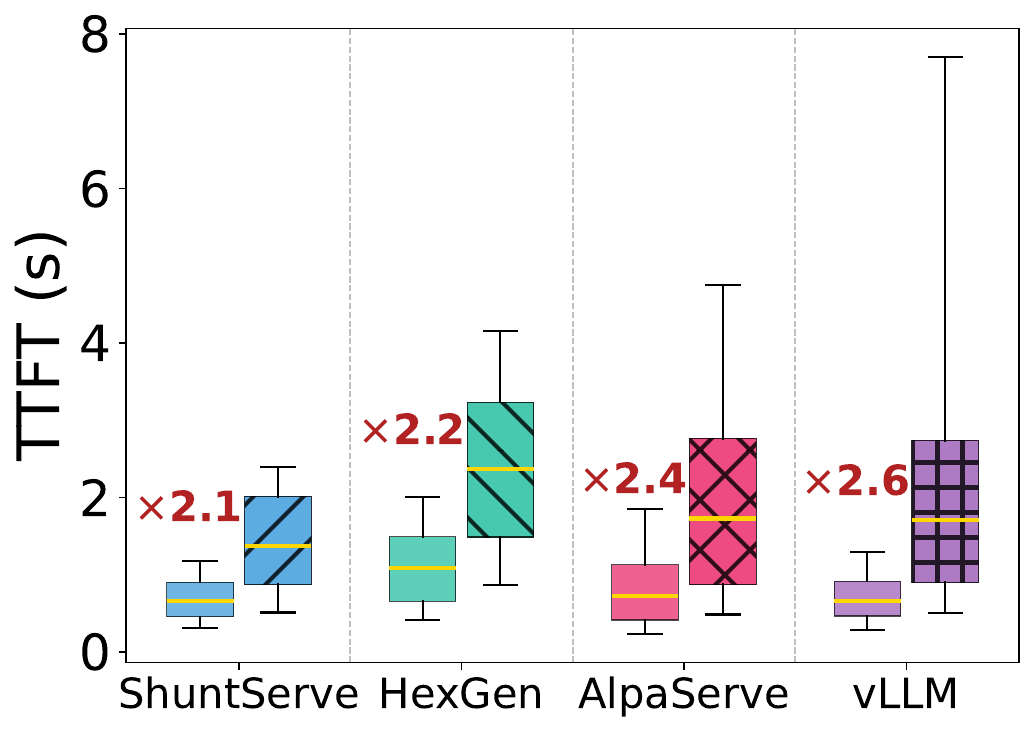}}
\hfill
\subfloat[TPOT\label{fig:model_placement_online_tpot}]{%
    \includegraphics[width=0.48\columnwidth]{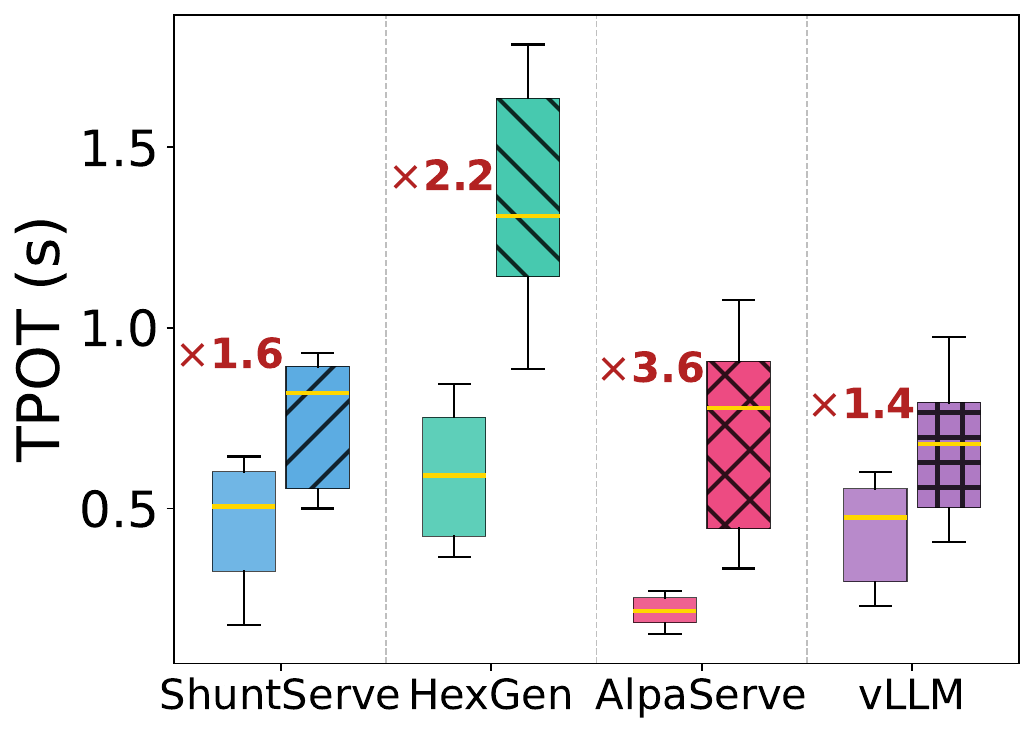}}
\caption{Online serving latency comparison under pipeline configurations determined by each baseline system's model placement algorithm. For each system, the solid box represents Qwen3-32B, while the hatched box represents Llama-3.1-70B.}
\label{fig:model_placement_online_latency}
\end{figure}

We next evaluate request latency under online workload to confirm that the throughput improvements do not sacrifice request latency. For Llama-3.1-70B, we replay the first 3 minutes of the trace over a 15-minute window at approximately 0.7 req/s, and for Qwen3-32B, the first 9 minutes over a 15-minute window at approximately 2.4 req/s. These arrival rates remain within the serving capacity of all baselines, ensuring a fair latency comparison without distortion from queuing delays. Figure~\ref{fig:model_placement_online_latency} presents the TTFT (Figure~\ref{fig:model_placement_online_ttft}) and TPOT (Figure~\ref{fig:model_placement_online_tpot}) distributions. In each box, the span represents the 25th to 75th percentile, the whiskers indicate the 10th and 90th percentile, and the yellow line denotes the median. The annotated multipliers denote the factor of increase in median latency when scaling from Qwen3-32B to Llama-3.1-70B for each system.

In these results, ShuntServe achieves a median TTFT of 1.37s and TPOT of 0.82s for Llama-3.1-70B, and 0.65s TTFT and 0.50s TPOT for Qwen3-32B. Compared to vLLM, which applies the intuitive even-partitioning placement, ShuntServe shows 19\% lower median TTFT and 22\% higher median TPOT for Llama-3.1-70B, and 2\% lower TTFT and 6\% higher TPOT for Qwen3-32B. At the P90 tail latency, ShuntServe reduces TTFT by 69\% for Llama-3.1-70B (2.39s vs 7.70s) while maintaining comparable TPOT (0.93s vs 0.97s). 

Among the baselines, HexGen exhibits the highest latency across both models. In particular, the median TPOT for Llama-3.1-70B reaches 1.31s, approximately 60\% higher than ShuntServe. This stems from HexGen's genetic algorithm, which favors expanding the PP dimension to equalize processing capability across stages. For instance, HexGen assigns the four L4 GPUs in a g6.12xlarge instance to four individual pipeline stages with TP1, whereas other baselines group them as a single stage with TP4. While this deep pipeline increases batch size and enables concurrent processing of multiple requests through micro-batching, it fails to exploit parallel computation through intra-node high-bandwidth interconnects, resulting in significant latency degradation.

AlpaServe achieves the lowest TPOT among all baselines for Qwen3-32B at 0.22s. AlpaServe selects the most aggressive replication among the baselines, which distributes requests across more pipelines with smaller batch sizes per pipeline. This is advantageous under online workloads that do not saturate the system. However, this strategy allocates more GPU memory to model weights rather than KV cache, which degrades throughput. Furthermore, as model size increases, the advantage of AlpaServe's placement diminishes. As shown by the annotated multipliers in Figure~\ref{fig:model_placement_online_latency}, ShuntServe's median TPOT increases by 1.6$\times$ when scaling from Qwen3-32B to Llama-3.1-70B, whereas AlpaServe's increases by 3.6$\times$. This is because larger models impose tighter memory capacity constraints that limit the degree of replication, reducing the effectiveness of AlpaServe's placement strategy. These results confirm that ShuntServe's throughput-oriented placement does not sacrifice request latency, while exhibiting stable scaling behavior as model size increases.

\subsubsection{Algorithm Execution Time of Beam Search}\label{eval:top_k_sensitivity_analysis}

\begin{figure}[t]
\hspace{5mm}\includegraphics[width=0.8\columnwidth]{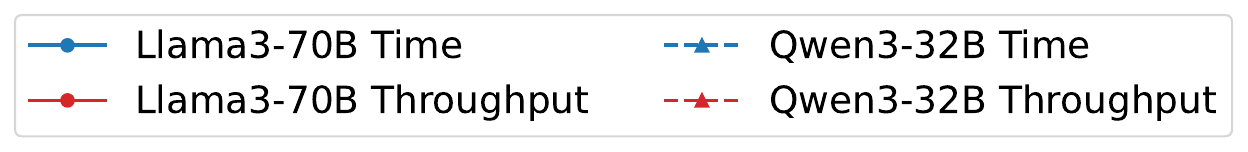}\\[-3mm]
\centering
\subfloat[24 GPU cluster (3 GPU types, 9 instances)\label{fig:top_k_dp_result_evaluation_cluster}]{%
    \includegraphics[width=0.48\columnwidth]{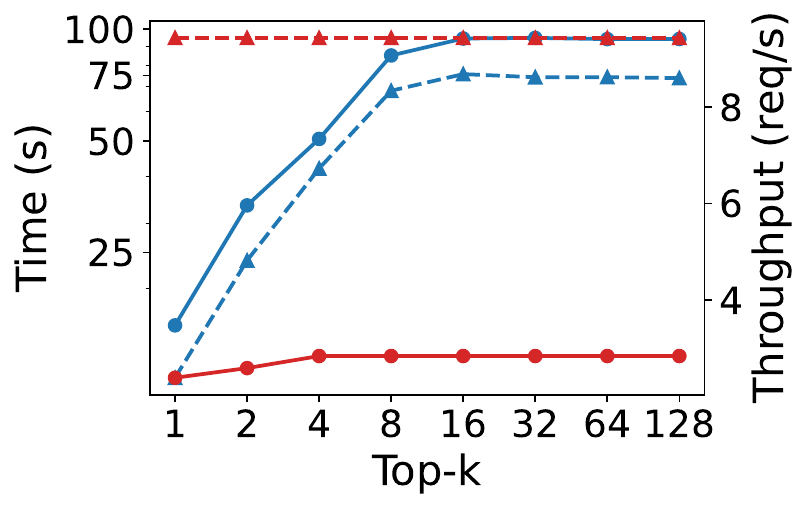}%
}
\hfill
\subfloat[76 GPU cluster (7 GPU types, 15 instances)\label{fig:top_k_dp_result_large_cluster}]{%
    \includegraphics[width=0.48\columnwidth]{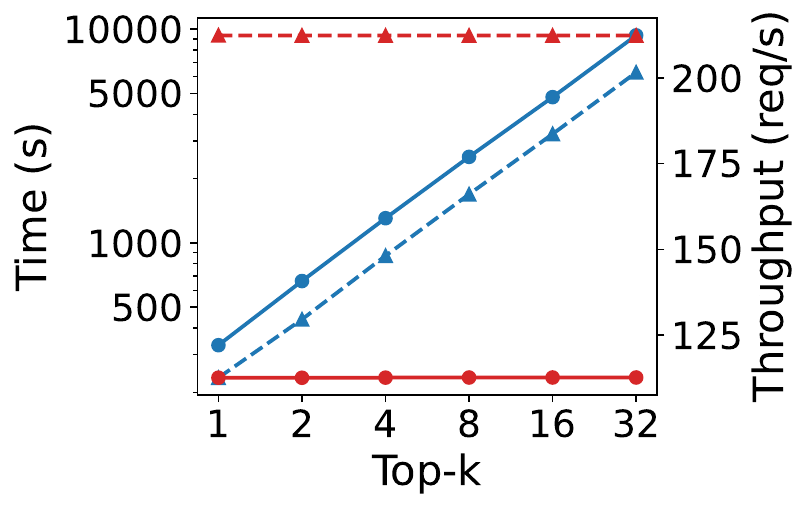}%
}
\caption{Effect of beam search width $k$ on algorithm execution time and the quality of the resulting model placement for Llama-3.1-70B and Qwen3-32B.}
\label{fig:top_k_dp}
\end{figure}

ShuntServe employs beam search to manage the combinatorially explosive search space of the DP-based optimizer, retaining only the top-$k$ candidate placements at each DP entry. This experiment evaluates the trade-off between the algorithm execution time and the quality of the resulting model placement as the beam size $k$ varies. Two cluster configurations are used: (1) the 24 GPU cluster, consisting of 3 GPU types across 9 instances, used in our main experiments, and (2) the 76 GPU cluster, consisting of 7 GPU types across 15 instances, configured with one instance of each of the 15 GPU instance types available on AWS. The algorithm execution time depends on the CPU used for optimization, with all measurements in this experiment conducted on an m8a.8xlarge instance.

Figure~\ref{fig:top_k_dp} presents the results, with the left y-axis showing the algorithm execution time on a log scale and the right y-axis showing the estimated throughput of the resulting placement. For the 24 GPU cluster (Figure~\ref{fig:top_k_dp_result_evaluation_cluster}), the algorithm completes in approximately 15.90s for Llama-3.1-70B and 11.49s for Qwen3-32B at $k=1$, increasing linearly up to $k=8$. Beyond $k=11$, execution time converges to approximately 94s for Llama-3.1-70B and 75s for Qwen3-32B, as the number of feasible candidate model placements in the 24 GPU cluster becomes smaller than $k$, eliminating the pruning of beam search. Figure~\ref{fig:top_k_dp_result_large_cluster} shows the results on the 76 GPU cluster. Execution time grows linearly with $k$ throughout the measured range up to $k=32$, reaching 9,433s for Llama-3.1-70B and 6,418s for Qwen3-32B, as the number of possible stage configurations at each DP entry increases combinatorially with the number of GPU types and instances in the cluster.

In terms of placement quality, the throughput of the resulting placement is largely insensitive to $k$. For Llama-3.1-70B on the 24 GPU cluster, throughput increases from 2.382 req/s at $k=1$ to 2.830 req/s at $k=3$, where it reaches a plateau. On the 76 GPU cluster, throughput varies by only approximately 0.1\% across all $k$ values. For Qwen3-32B, the optimizer finds identical placements across all $k$ values on the 24 GPU cluster at 9.424 req/s, and exhibits similarly negligible variation on the 76 GPU cluster. These results indicate that the DP-based optimizer of ShuntServe effectively identifies near-optimal placements even at small $k$ values, as the partial placements in the DP table closely satisfy the optimal substructure property. Based on these findings, we set $k=3$ in our main experiments to balance algorithm execution time and placement quality.

\subsubsection{One-time Calibration Overhead}\label{eval:calibration_overhead}

\begin{table}[t]
\centering
\small
\caption{Calibration overhead}
\label{tab:calibration_overhead}
\begin{tabular}{llcr}
\toprule
\textbf{Instance} & \textbf{GPU} & \textbf{Batch Sizes} & \textbf{Time (s)} \\
\midrule
g5.48xlarge   & NVIDIA A10G  & [1, 2, 4, \ldots, 128] & 317.98 \\
g6.48xlarge   & NVIDIA L4    & [1, 2, 4, \ldots, 128] & 204.50 \\
g6e.48xlarge  & NVIDIA L40S  & [1, 2, 4, \ldots, 512] & 499.88 \\
\midrule
\textbf{Total} & --           & --                     & \textbf{1022.36} \\
\bottomrule
\end{tabular}
\end{table}

We initially pursued a fully profiling-free performance estimation by relying on the official hardware white papers; however, we observed substantial discrepancies between the reported and actual performance. For instance, the L4 GPU's official FP16 tensor core FLOPS is reported as 121 TFLOPS (non-sparse), whereas our measurements yielded approximately 55 TFLOPS. Calibration is therefore essential for determining the effective performance of the underlying hardware.

Table~\ref{tab:calibration_overhead} summarizes the calibration time for each GPU type used in the evaluation of ShuntServe. We represent each hardware feature as a single unified scalar (i.e., effective FLOPS, memory bandwidth, and network bandwidth), which is invariant to the serving configuration. To obtain a representative scalar for each feature, we measure the feature at multiple batch sizes and take the median of the measured values. For all GPU instances, we selected the largest available size of the family, 48xlarge (i.e., eight GPUs per instance), and reused the resulting calibration data for smaller instances within the same family.

In total, the calibration required for our experiments consumed 1022.36 seconds, corresponding to 0.28 GPU-hours. Relative to the approximately 1,000 GPU-hours consumed by the full set of experiments, this overhead accounts for 0.03\%, which is negligible.

\subsection{Enhanced Tolerance to Spot Interruptions}\label{eval:interruption_tolerance}

\begin{figure}[t]
\centering
\includegraphics[width=0.65\columnwidth]{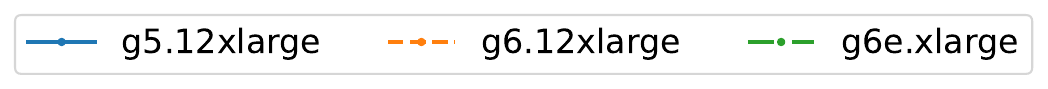}\\[-0mm]
\includegraphics[width=0.65\columnwidth]{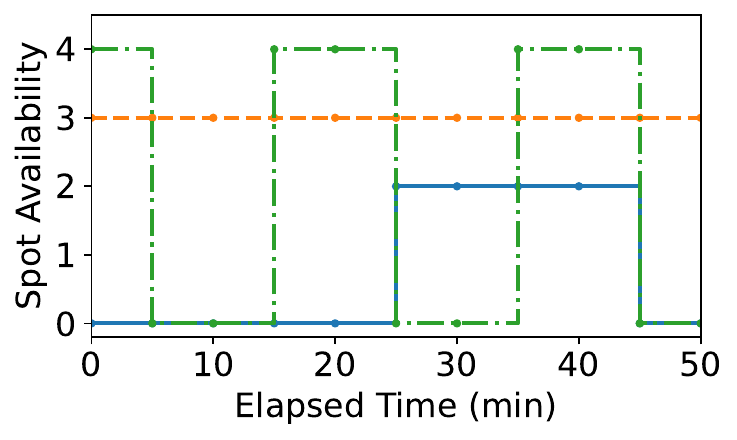}
\caption{Spot instance availability over time for three GPU instance types under the evaluation scenario.}
\label{fig:spot_availability_scenario}
\end{figure}

We evaluate ShuntServe's fault tolerance mechanisms for handling spot instance interruptions, namely output-preserving request migration and concurrent initialization. To ensure consistent evaluation across all experiments, we extract a 50-minute spot availability scenario by tracking the availability of each instance type and simulating the corresponding spot events. Figure~\ref{fig:spot_availability_scenario} illustrates this scenario. The scenario is extracted from a 6-day trace spanning March 13--18, 2026, which corresponds to the same observation period as Figure~\ref{fig:spot_availability}. To select the most challenging scenario, we assign each candidate window a composite score based on the frequency of availability change events and the magnitude of instances affected by each event, and select the window with the highest score. This ensures that the evaluation reflects worst-case conditions among the observed traces. We further observe that approximately 40.4\% (688 of 1,701) of the candidate trace windows have a score of zero, indicating no availability changes. In such scenarios, using spot instances directly reduces cost without any interruption overhead.

\paragraph{Baselines} To the best of the authors' knowledge, ShuntServe is the first system to leverage spot instances for LLM serving on heterogeneous GPU clusters. Although SpotServe~\cite{spotserve} addresses LLM serving on spot instances, it targets homogeneous GPU clusters, and its core fault tolerance mechanism based on the Kuhn-Munkres algorithm for device mapping does not generalize to heterogeneous pipeline configurations. SkyServe~\cite{skyserve-multi-region-inference} addresses spot instance utilization for AI serving in cloud environments. Yet, it considers higher-level concerns such as instance provisioning, placement, and scheduling, not the fault tolerance of distributed LLM serving. We therefore evaluate ShuntServe's spot tolerance against the following baselines. \emph{On-demand Only} uses exclusively on-demand instances without any interruptions. \emph{No Handle} uses spot instances but applies no fault tolerance mechanism upon interruption. \emph{Request Migration} implements output-preserving request migration on vLLM without concurrent initialization. \emph{Concurrent Initialization} implements concurrent initialization on vLLM without request migration. ShuntServe combines both mechanisms.

\subsubsection{Serving Throughput under Spot Interruptions}

\begin{figure}[t]
\centering
\includegraphics[width=0.95\columnwidth]{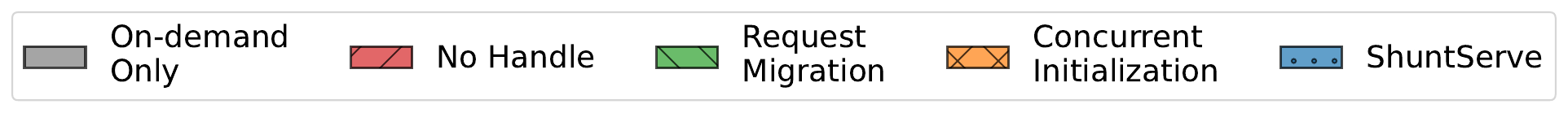}\\[-3mm]
\subfloat[Llama-3.1-70B\label{fig:fault_tolerance_llama_throughput}]{%
    \includegraphics[width=0.48\columnwidth]{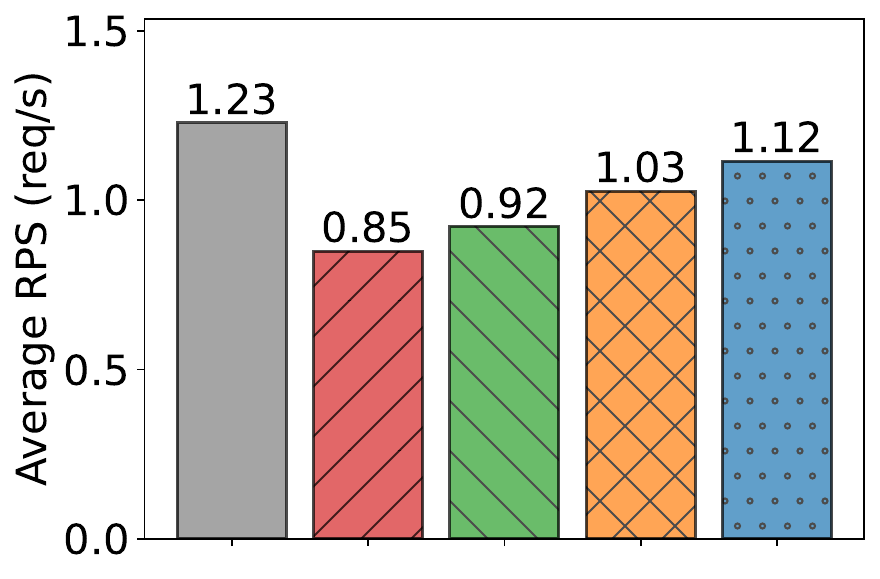}}
\hfill
\subfloat[Qwen3-32B\label{fig:fault_tolerance_qwen_throughput}]{%
    \includegraphics[width=0.48\columnwidth]{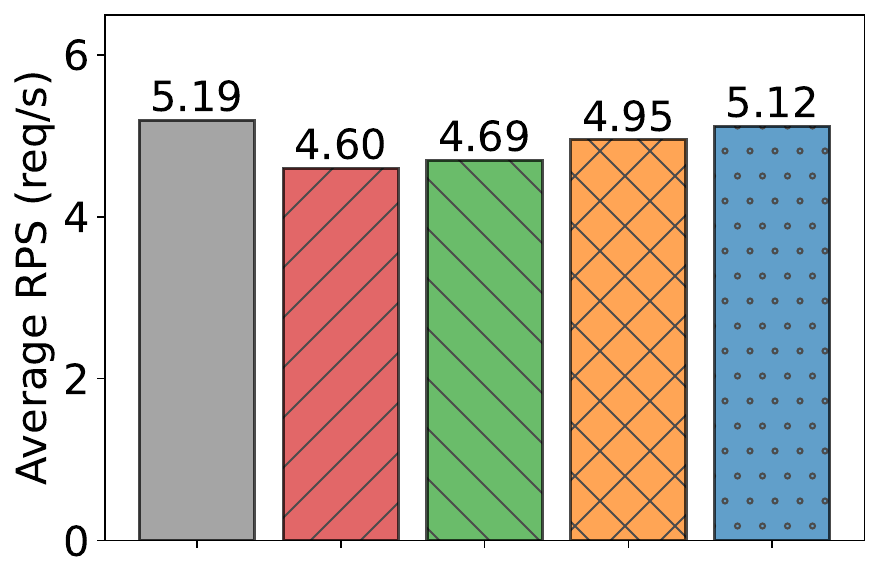}}
\caption{Offline serving throughput under a spot availability scenario. The y-axis represents throughput in RPS (req/s). (a) shows results for Llama-3.1-70B, and (b) shows results for Qwen3-32B.}
\label{fig:fault_tolerance_throughput}
\end{figure}

We evaluate the impact of spot interruptions on offline workload. The entire dataset is dispatched to the system at once. After a 50-minute spot availability window, the system is halted and the serving metrics are collected from the requests completed within the window. Figure~\ref{fig:fault_tolerance_throughput} presents the results.

For Llama-3.1-70B (Figure~\ref{fig:fault_tolerance_llama_throughput}), as expected, On-demand Only achieves the highest throughput at 1.23 req/s since it is not affected by spot interruptions, while No Handle yields the lowest at 0.85 req/s. Request Migration improves throughput by approximately 8\% over No Handle, reaching 0.92 req/s, by recovering the computation of in-flight requests that would otherwise be discarded upon interruption. Concurrent Initialization yields a larger improvement of approximately 21\% to 1.03 req/s by reducing the pipeline downtime through overlapping replacement node preparation with ongoing serving. ShuntServe, which combines both mechanisms, achieves 1.12 req/s, a 32\% improvement over No Handle and the highest throughput among the baselines that utilize spot instances.

For Qwen3-32B (Figure~\ref{fig:fault_tolerance_qwen_throughput}), the results follow a similar trend, but the gap between the baselines that utilize spot instances and On-demand Only narrows considerably. Specifically, the gap between On-demand Only and No Handle decreases from approximately 45\% for Llama-3.1-70B to approximately 13\% for Qwen3-32B (i.e., 5.19 req/s vs 4.60 req/s). This difference arises from the characteristic of distributed LLM serving where the interruption of a single instance causes the entire pipeline to stall. Larger models require pipelines with more nodes, making them more vulnerable to this cascading effect. Conversely, smaller models require fewer nodes per pipeline, resulting in more independent pipelines within the cluster and reducing the scope of each interruption. For example, ShuntServe constructs 2 pipelines for Llama-3.1-70B and 4 pipelines for Qwen3-32B on the evaluation cluster. This indicates that leveraging spot instances is particularly effective for serving smaller models. Consequently, ShuntServe achieves 5.12 req/s for Qwen3-32B, with only approximately 1\% throughput difference compared to On-demand Only.

\subsubsection{Temporal Analysis of Latency under Spot Interruptions}

\begin{figure*}[t]
\centering
\includegraphics[width=0.7\textwidth]{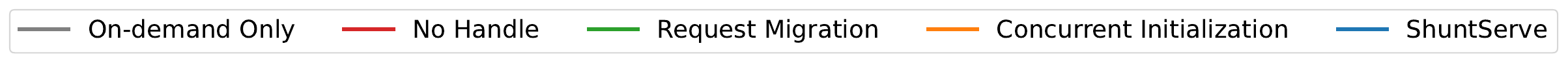}\\[-3mm]
% Row 1: Llama-3.1-70B
\subfloat[Llama-3.1-70B end-to-end latency (Mean)\label{fig:fault_tolerance_llama_avg}]{%
    \includegraphics[width=0.48\textwidth]{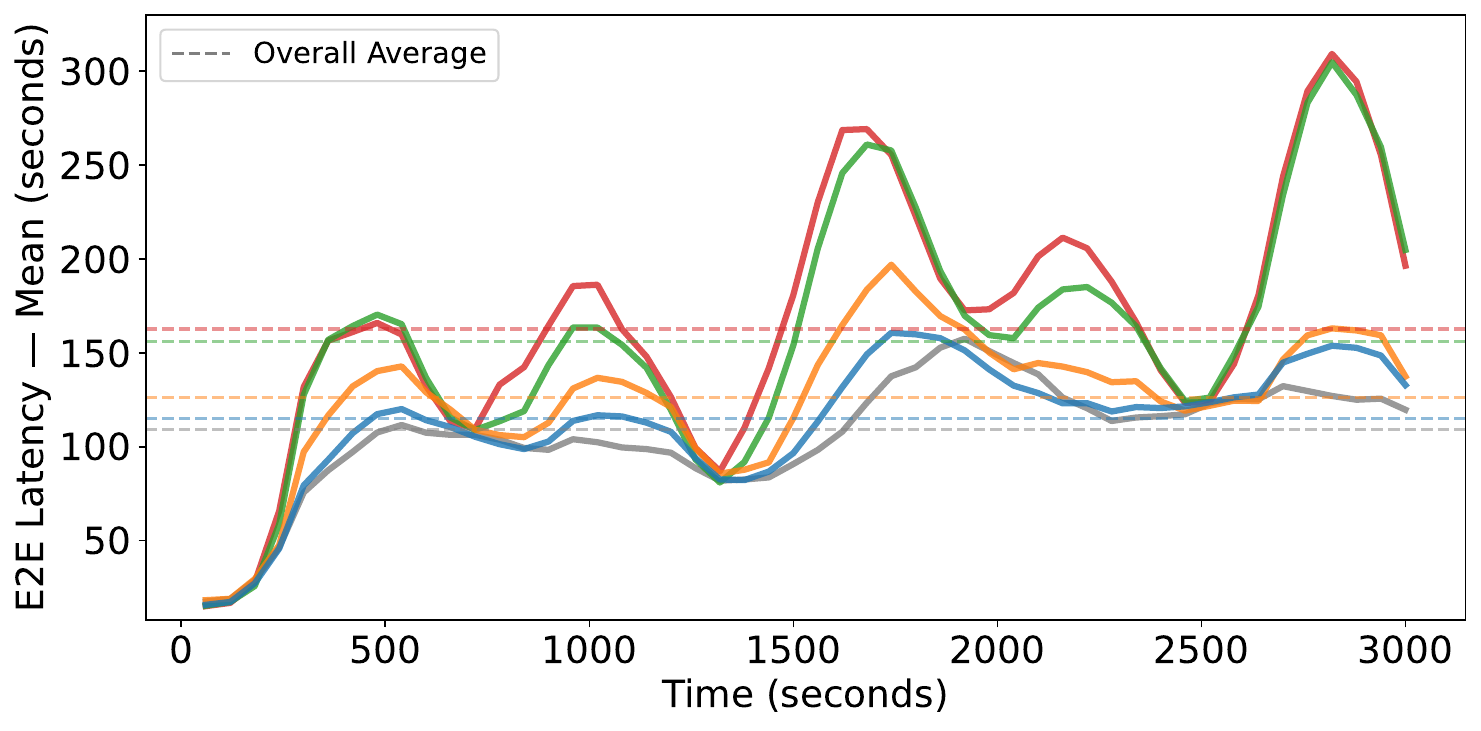}}
\hfill
\subfloat[Llama-3.1-70B end-to-end latency (P90)\label{fig:fault_tolerance_llama_p90}]{%
    \includegraphics[width=0.48\textwidth]{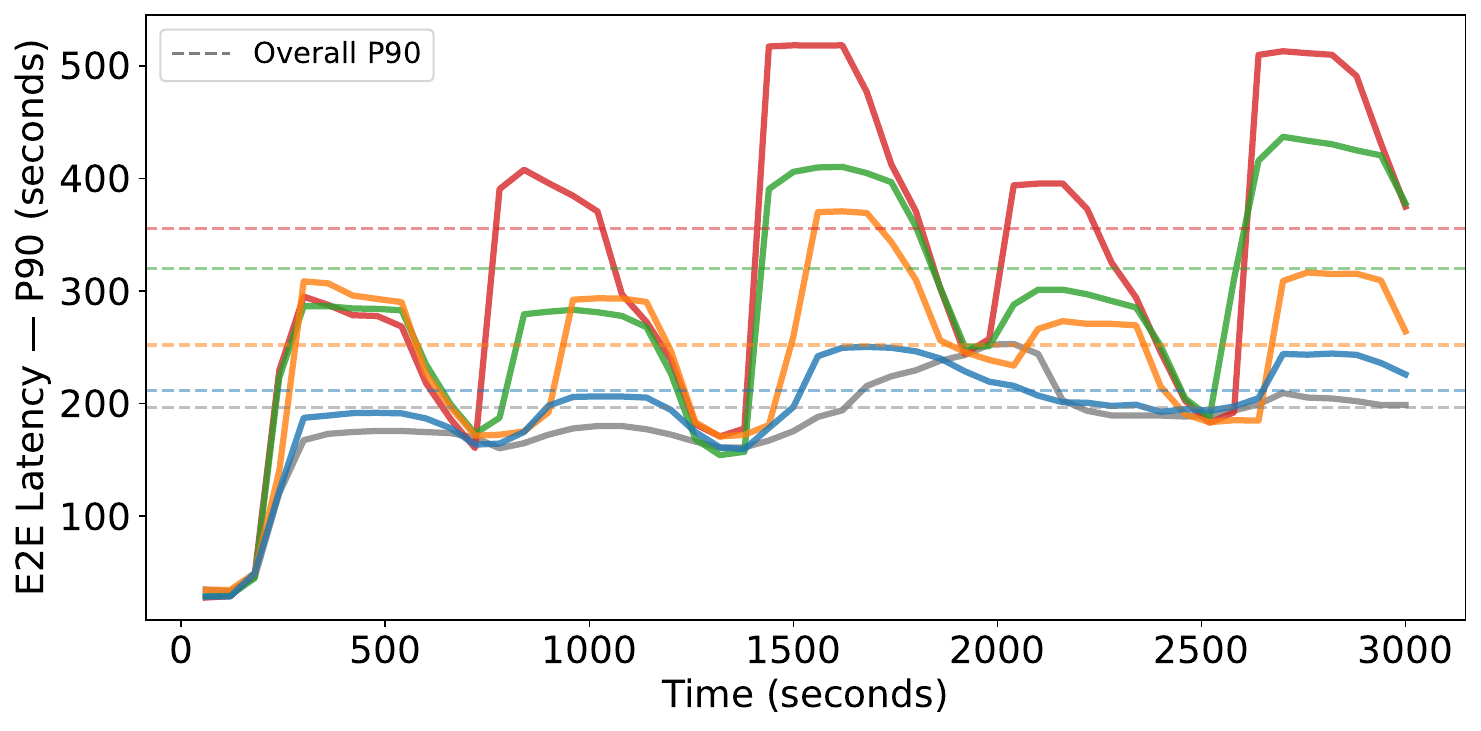}}\\
\vspace{-2mm}
% Row 2: Qwen3-32B
\subfloat[Qwen3-32B end-to-end latency (Mean)\label{fig:fault_tolerance_qwen_avg}]{%
    \includegraphics[width=0.48\textwidth]{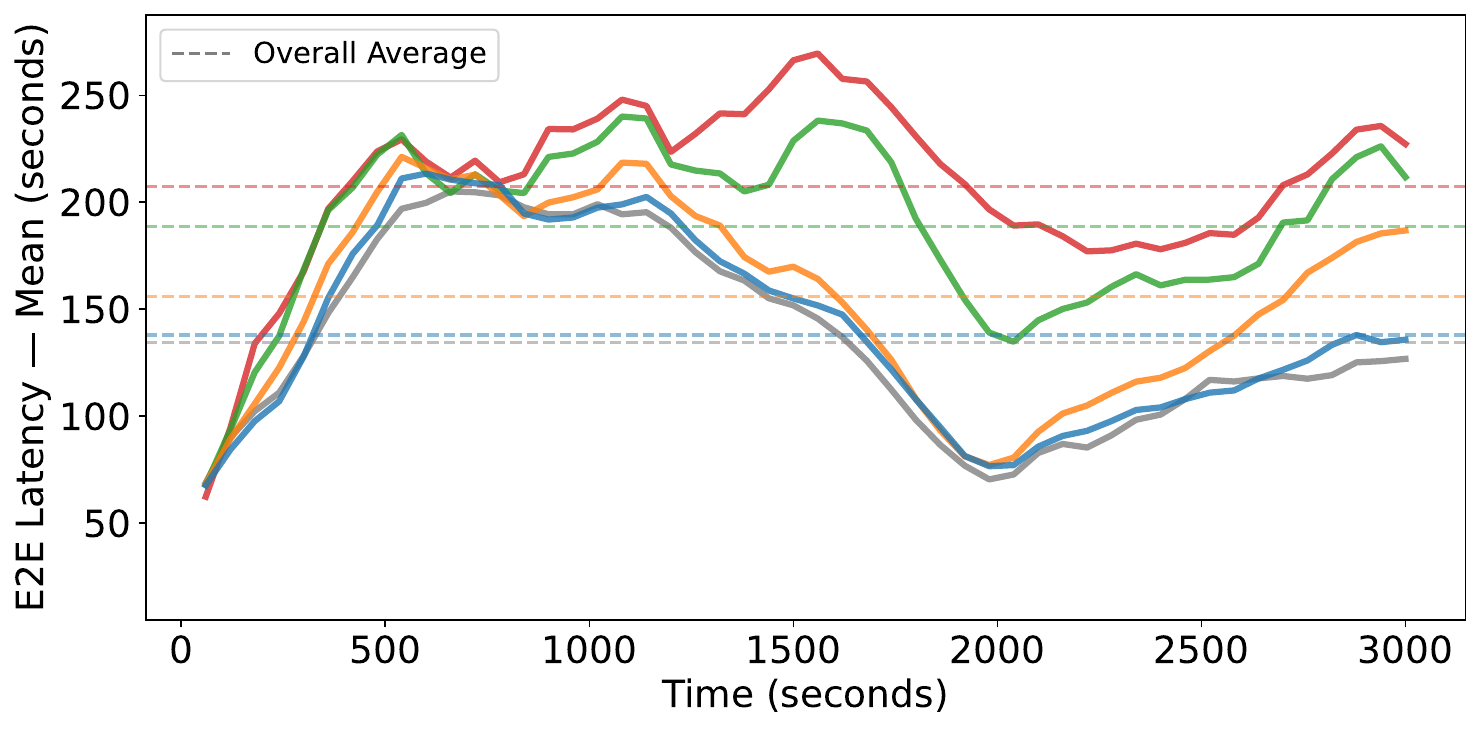}}
\hfill
\subfloat[Qwen3-32B end-to-end latency (P90)\label{fig:fault_tolerance_qwen_p90}]{%
    \includegraphics[width=0.48\textwidth]{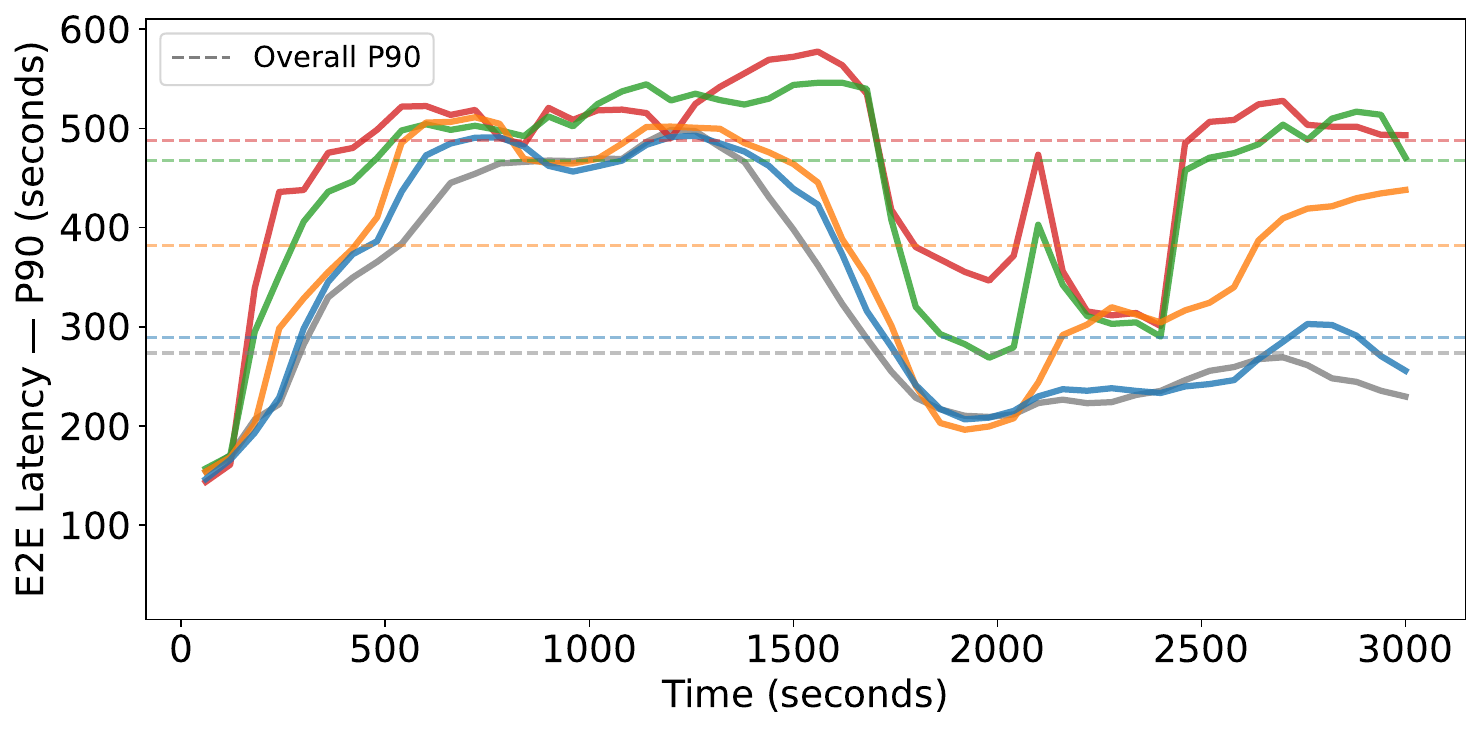}}
\caption{End-to-end latency trends over time under the spot availability scenario. Each data point represents a 5-minute trailing moving average of end-to-end request latency. (a) and (b) show results for Llama-3.1-70B, and (c) and (d) show results for Qwen3-32B.}
\label{fig:ft_timeline}
\end{figure*}

We further analyze the impact of spot interruptions on real-time request processing by measuring the temporal evolution of end-to-end latency under online workloads. We evaluate both mean and P90 end-to-end latency to capture the average behavior and the tail behavior, respectively. As in the online evaluation of Section~\ref{eval:online_model_placement}, the original arrival pattern is preserved while the overall request rate is scaled to keep all baselines below saturation. For Llama-3.1-70B, the arrival intervals are scaled by a factor of six, while for Qwen3-32B, the original trace is replayed unmodified, as the system throughput exceeds the original arrival rate. Figure~\ref{fig:ft_timeline} presents the temporal trends of mean and P90 end-to-end latency for both models across the five baselines, with the x-axis representing elapsed time in seconds and the y-axis representing end-to-end latency in seconds. Each data point represents a 5-minute trailing moving average of end-to-end request latency, and horizontal dashed lines indicate the overall metric for each baseline across the entire serving window.

For Llama-3.1-70B, No Handle exhibits the highest overall mean latency at 162.67s in Figure~\ref{fig:fault_tolerance_llama_avg}. Tail latency is even more severely affected, with each baseline exhibiting pronounced peaks in Figure~\ref{fig:fault_tolerance_llama_p90} in response to spot interruption events. For instance, the spot interruption at approximately 25 minutes drives the temporal P90 latency of No Handle beyond 500s. As a result, its overall P90 latency reaches 355.51s, approximately 81\% higher than the 196.39s of On-demand Only. Among the baselines that incorporate fault tolerance, Request Migration achieves a mean latency of 155.88s and a P90 latency of 320.27s, while Concurrent Initialization achieves 126.10s and 251.95s, respectively.

For Qwen3-32B, where each pipeline consists of fewer instances and the cluster hosts a larger number of pipelines, the baselines exhibit a similar pattern to Llama-3.1-70B in Figures~\ref{fig:fault_tolerance_qwen_avg} and~\ref{fig:fault_tolerance_qwen_p90}. No Handle exhibits the highest mean and P90 latencies at 207.43s and 487.87s, followed by Request Migration at 188.70s and 467.86s, and Concurrent Initialization at 156.01s and 382.15s.

Across both models, Concurrent Initialization consistently achieves lower mean and P90 latency than Request Migration. This gap stems from the limitation of output-preserving request migration, which preserves the state of in-flight requests upon interruption but does not reduce the downtime itself. When a pipeline becomes unavailable during the downtime, incoming requests are concentrated on the remaining pipelines, increasing their batch sizes and inducing additional queueing delay. The resulting latency degradation persists until the affected pipeline is restored and the accumulated backlog is fully processed. These results indicate that minimizing downtime is critical for latency-sensitive workloads such as online serving. Consequently, ShuntServe, which combines both mechanisms, achieves the lowest latency among all baselines on both models, with mean and P90 latencies of 115.08s and 211.44s on Llama-3.1-70B and 138.00s and 289.09s on Qwen3-32B, corresponding to only 2.6\%--7.7\% increases over On-demand Only.

\subsubsection{Cost Efficiency Comparison}\label{eval:cost_efficiency}

\begin{figure}[t]
\includegraphics[width=\columnwidth]{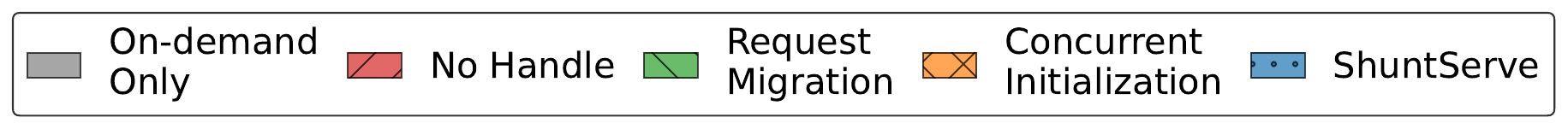}
\centering
\subfloat[Cost efficiency on Llama-3.1-70B.\label{fig:cost_efficiency_comparison_llama3}]{%
    \includegraphics[width=0.48\columnwidth]{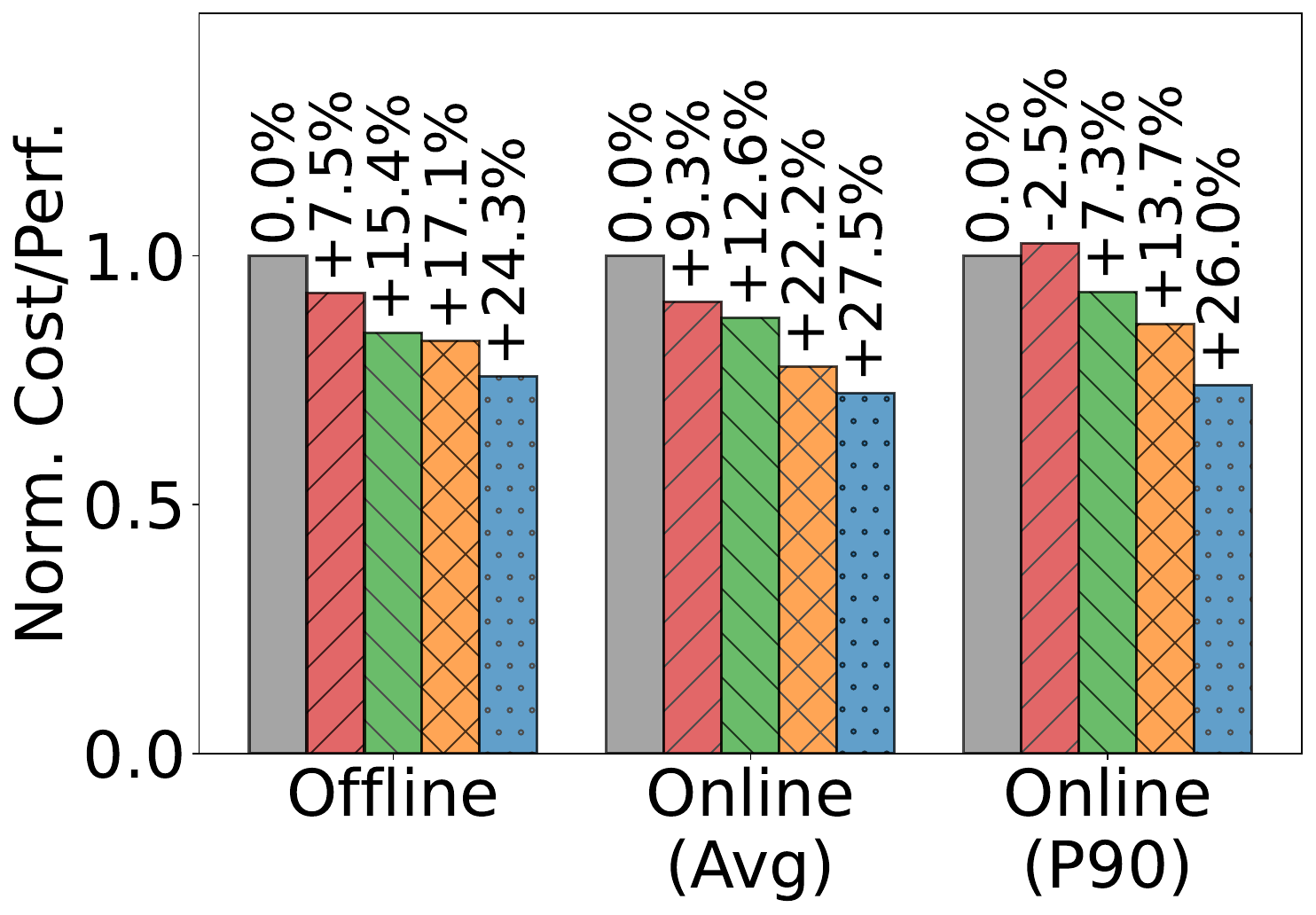}}
\hfill
\subfloat[Cost efficiency on Qwen3-32B.\label{fig:cost_efficiency_comparison_qwen3}]{%
    \includegraphics[width=0.48\columnwidth]{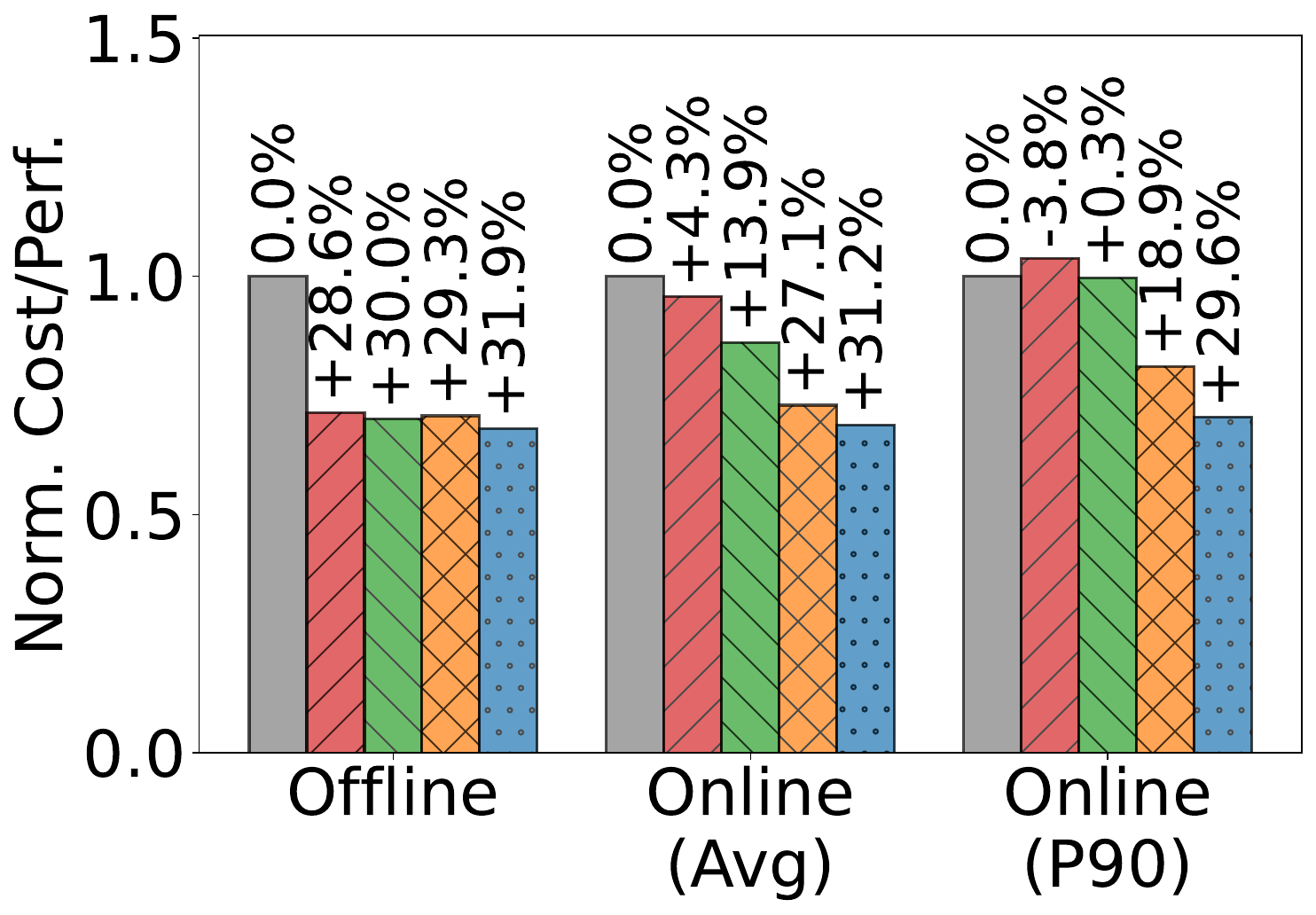}}
\caption{Cost efficiency comparison across baselines, normalized to On-demand Only (lower is better). The percentage above each bar denotes the reduction relative to On-demand Only.}
\label{fig:cost_comparison_efficiency}
\end{figure}

The operating cost over the 50-minute window amounts to \$27.16 for On-demand Only, \$17.05 for Request Migration and No Handle, and \$18.15 for Concurrent Initialization and ShuntServe. Concurrent Initialization and ShuntServe incur approximately \$1.10 more than the other spot-based baselines because the interrupted node remains active until the new pipeline initialization completes, resulting in temporary overlap of billing between the interrupted and replacement nodes. To evaluate whether these cost reductions translate into actual cost efficiency gains after accounting for the performance degradation caused by spot interruptions, Figure~\ref{fig:cost_comparison_efficiency} presents the cost efficiency of each baseline normalized to On-demand Only. Even No Handle achieves lower cost per performance than On-demand Only in most cases, as the cost savings from spot pricing outweigh the performance loss. The exception arises in the P90 tail latency, where the impact of spot interruptions is most pronounced. For Llama-3.1-70B, No Handle exhibits a 2.5\% higher cost per performance than On-demand Only.

For Llama-3.1-70B, Concurrent Initialization achieves lower cost per performance than Request Migration across all three workload settings (i.e., offline, online mean, and online P90) in Figure~\ref{fig:cost_efficiency_comparison_llama3}. This stems from the fact that larger models occupy a greater portion of the cluster within each pipeline, expanding the scope affected by an interruption and amplifying the benefit of downtime reduction. In contrast, Qwen3-32B distributes the cluster across more pipelines, dispersing the impact of a single interruption. As a result, Request Migration achieves slightly lower cost per performance than Concurrent Initialization in the offline workload (Figure~\ref{fig:cost_efficiency_comparison_qwen3}). For online workloads, however, downtime affects request latency more severely, and Concurrent Initialization achieves lower cost per performance than Request Migration in both online mean and P90 metrics. ShuntServe, which combines both mechanisms, achieves the lowest cost per performance across all models and workloads, with cost efficiency improvements of up to 31.9\% over On-demand Only.

\subsubsection{Concurrent Initialization Overhead}\label{eval:concurrent_initialization}

\begin{figure}[t]
\centering
\subfloat[Concurrent initialization timeline.\label{fig:node_replacement_overhead_flow}]{%
    \includegraphics[width=0.70\columnwidth]{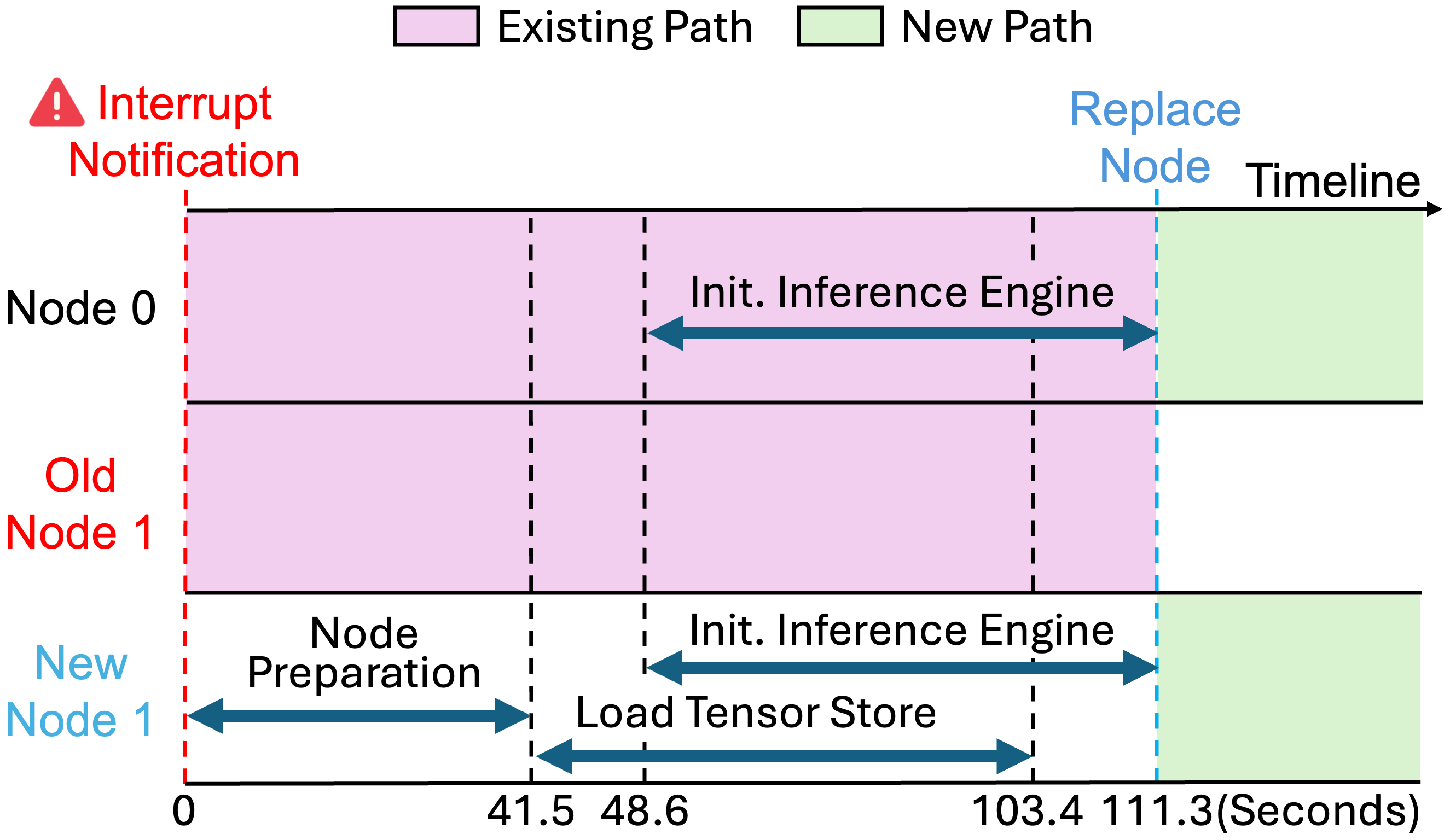}}
\subfloat[Time breakdown by component.\label{fig:node_replacement_overhead_graph}]{%
    \includegraphics[width=0.26\columnwidth]{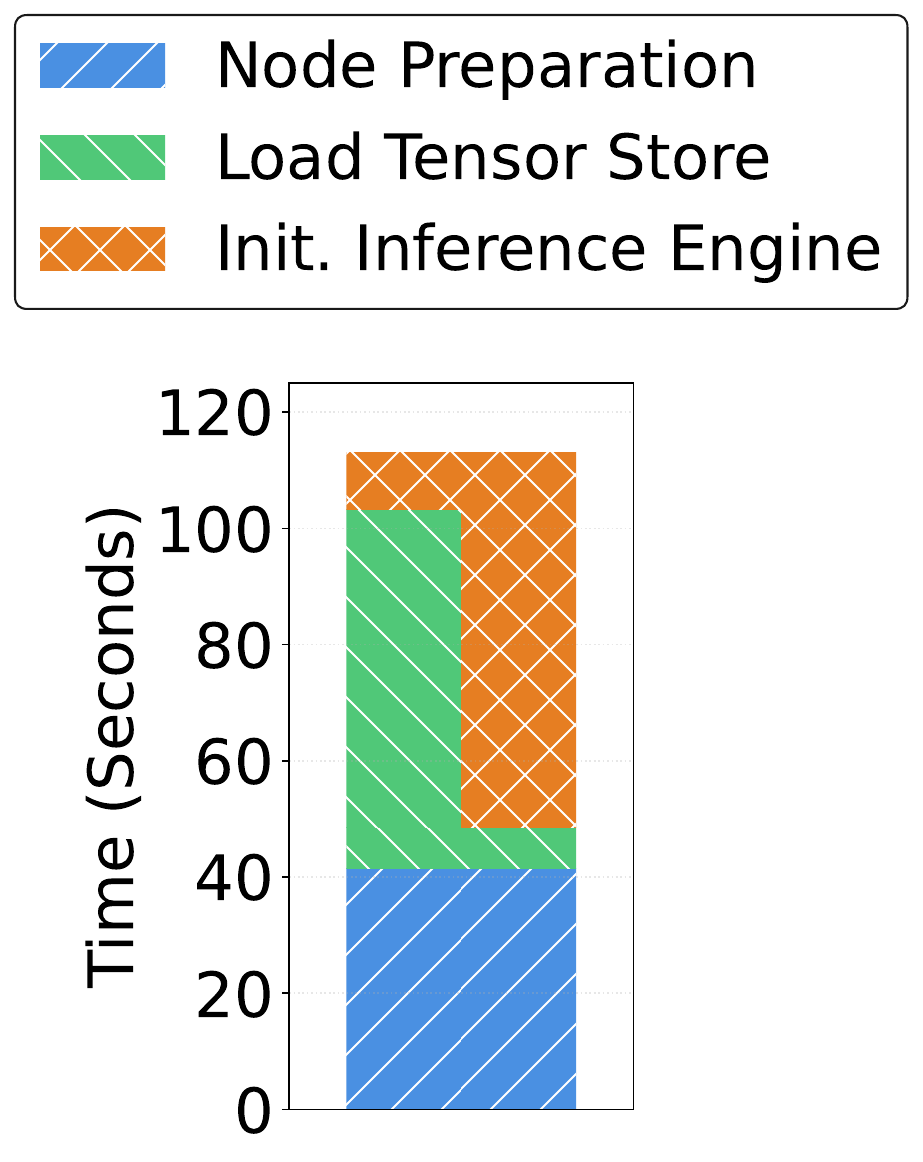}}
\caption{Concurrent initialization overhead analysis, averaged across the measured runs.}
\label{fig:node_replacement_overhead}
\end{figure}

We present a time breakdown across the components of the concurrent initialization process to quantify the duration of new pipeline preparation upon spot interruption. The node preparation time is measured six times for each of the three instance types in the evaluation cluster, yielding 18 measurements in total. The Shared Tensor Store and Inference Engine loading times are measured from the completion of node preparation until both components complete loading, repeated 10 times across both pipelines for Llama-3.1-70B, which incurs a longer Shared Tensor Store loading time than Qwen3-32B. Figure~\ref{fig:node_replacement_overhead} presents the results.

The replacement node was provisioned and became ready for inference in 41.55s on average with a standard deviation of 7.54s. Once the node was ready, the Shared Tensor Store process was launched on it, loading the model partition from remote storage in 61.85s on average (std 9.59s, max 81.45s). In parallel, a new Inference Engine was initialized on both the unaffected node (i.e., Node 0) and the replacement node (i.e., New Node 1), taking 64.51s on average (std 9.25s, max 81.84s). The total initialization time amounts to approximately 111.3s on average. The downtime that ShuntServe incurs corresponds to the portion of the initialization time that exceeds the grace period. Since the grace period varies across cloud providers from 30s in Azure to 120s in AWS to 300s in Alibaba Cloud, the resulting downtime varies accordingly. In the AWS environment used in our evaluation, the average duration falls within the 120s grace period, allowing ShuntServe to achieve near-zero downtime upon spot interruption.

\section{Discussion}\label{sec:discussion}

The evaluation results demonstrate that ShuntServe achieves substantial cost efficiency on heterogeneous spot GPU clusters while maintaining comparable service quality to on-demand baselines. We discuss the limitations of the current design in Section~\ref{subsec:limitation} and outline directions for future extension in Section~\ref{subsec:future_work}.

\subsection{Limitation}\label{subsec:limitation}

\paragraph{Fidelity of the Estimator under Real-World Workloads} 
Optimization algorithms for identifying the best model placement must evaluate and compare candidate placements throughout their search process; accurate prediction of the serving performance of a given placement configuration is therefore essential. ShuntServe achieves an average MAPE of 6.63\% through roofline-based static latency estimation and throughput modeling, yet this accuracy is established under a synthetic dataset with fixed input/output lengths and a static batching environment. 

In real-world workloads, however, context length distributions are highly heterogeneous and request arrival patterns are dynamic. Moreover, the analytical formulation does not adequately capture non-linear factors such as kernel launch overhead, SM utilization, and cache effects. Prior systems such as Vidur~\cite{vidur} and Habitat~\cite{habitat} adopt ML-based approaches to model these non-linear characteristics, but at the cost of profiling that is sensitive to the combinatorially large configuration space. 

A promising direction for future work is to combine the strengths of these approaches. Integrating ShuntServe's analytical formulation with an ML-based approach could reduce the profiling burden while capturing non-linear effects, and further integration with a simulation-based approach could enable adaptation to dynamic workload regimes.

\paragraph{Fault Tolerance for Long Context Workload}
While the recomputation-based approach is sufficiently efficient for restoring the state of in-flight requests, it can incur higher overhead than KV cache transfer under very long context lengths (e.g., in the comparative measurements of Section~\ref{subsubsec:request_migration}, recomputation on L40S exhibits approximately 9.6\% higher latency than transfer at a 64K context length). This overhead stems from the quadratic cost of computing attention scores in transformer layers, and can result in considerable tail latency degradation for workloads characterized by very long contexts. KV cache transfer, however, has a strict constraint in spot environments that the transfer, whose duration depends on the data size and the available network bandwidth, must complete within the grace period, making it unsuitable as a sole request recovery mechanism. 

A promising direction for future work is thus a hybrid recovery scheme that tracks the progress of in-flight requests and the remaining grace period, and selects an appropriate request recovery mechanism for each request individually. For instance, cache transfer may be elected for requests with very long contexts, while recomputation is retained for shorter ones.

\subsection{Future Work}\label{subsec:future_work}

\paragraph{Adapting to Spot Volatility across Time and Regions}
ShuntServe currently optimizes model placement over a static heterogeneous GPU cluster and responds to interruptions through auxiliary on-demand fallback. Spot environments, however, exhibit substantial fluctuations across both time and regions. 

Along the temporal dimension, the system could be extended to support the flexible membership of heterogeneous GPU instances in a manner that maximizes spot availability within the cluster, which would require a model placement algorithm capable of rapidly reconfiguring the pipeline topology in response to the dynamically changing cluster. 

Furthermore, while the current design assumes operation within a single region, extending ShuntServe to multiple regions could yield additional cost savings and motivates a multi-region model placement optimizer tailored to the parallelization characteristics of LLM serving pipelines (e.g., replication incurs little traffic across replicas and is therefore well suited to placement across regions~\cite{skyserve-multi-region-inference}).

\paragraph{Phase-Aware Model Placement}
Prefill-decode disaggregation techniques~\cite{distserve,splitwise,thunderserve,pmlr-v267-jiang25c} can be integrated to maximize resource utilization in clusters composed of heterogeneous GPU instances. The distinct resource requirements of the prefill and decode phases in generative LLM inference expand the optimization space for model placement. Realizing this direction requires per-phase latency prediction and throughput modeling, as well as techniques for mitigating the KV cache transfer bottleneck that arises across heterogeneous instances, particularly under limited network bandwidth.

\section{Related Work}\label{sec:related_work}

\begin{table*}[t]
\caption{Comparison with existing LLM inference systems}
\label{tab:related_work_comparison}
\centering
\begin{tabular}{lccccc}
\toprule
 & Vidur & AlpaServe & SpotServe & HexGen & ShuntServe \\
\midrule
Configuration-free Profiling      & - & - & - & \checkmark & \checkmark \\
Heterogeneous GPU Pipeline  & - & - & - & \checkmark & \checkmark \\
Model Placement Algorithm   & Exhaustive & DP+Beam Search & Exhaustive & DP+Genetic & DP+Beam Search \\
Interruption Handling       & - & - & \checkmark & - & \checkmark \\
\bottomrule
\end{tabular}
\end{table*}
\paragraph{Machine Learning System on Spot Instances}
Several studies have investigated predicting and handling spot instance interruptions to achieve cost-efficient training and serving of machine learning models in cloud environments. Bamboo \cite{bamboo-nsdi} provides fault-tolerant DNN training via redundant forward and backward pass computations across neighboring pipeline nodes. Parcae \cite{parcae-nsdi} predicts spot interruptions and proactively migrates pipelines to optimize liveput. SpotServe \cite{spotserve} pioneered distributed LLM serving on spot instances using dynamic re-parallelization and KV cache migration.

\paragraph{LLM Inference Latency Prediction on Various GPUs} Among existing LLM inference performance estimation approaches, Vidur~\cite{vidur} reports latency prediction errors below 9\%, but requires extensive operator-level profiling across a combinatorially explosive configuration space that varies with models, GPU types, parallelization strategies, input/output sequence lengths, and batch sizes. MaverIQ~\cite{maverIQ} reduces profiling cost by introducing a \emph{lightweight fingerprint} that contains only a few unique layers of the target LLM, and reports 1.3--1.7$\times$ lower estimation error than Vidur. However, MaverIQ still requires profiling across the configuration space, which does not fundamentally resolve the scalability challenge of profiling-based approaches.

In contrast, ShuntServe's estimator requires only a one-time, lightweight hardware calibration (FLOPS, memory bandwidth, and network bandwidth) for each GPU type, with negligible cost. The achieved MAPE of 6.63\% is competitive with these profiling-based approaches and demonstrates sufficient accuracy as the foundational component of the model placement optimizer.

\paragraph{LLM Serving System on Heterogeneous GPUs}
To address the scarcity of high-end GPUs, several studies have investigated maximizing cluster utilization by incorporating abundantly available commodity GPUs and accounting for the diverse performance characteristics of heterogeneous devices. Helix~\cite{helix} models throughput as a network flow and optimizes model placement using max-flow and MILP algorithms. HEXGEN~\cite{hexgen} models inference and network costs, optimizing placement via genetic algorithms and DP. Thunderserve~\cite{thunderserve} optimizes disaggregated serving by specializing heterogeneous devices based on their compute-bound and memory-bound characteristics.

Deploying LLM serving systems on heterogeneous spot GPU clusters introduces challenges that are not addressed by any single line of prior work. Systems designed for spot instances~\cite{spotserve,bamboo-nsdi,parcae-nsdi} assume homogeneous GPU clusters, which are vulnerable to correlated failures where all instances of the same type are simultaneously interrupted. Systems designed for heterogeneous clusters~\cite{thunderserve,helix,hexgen} do not target spot instance environments and therefore lack fault tolerance mechanisms. For instance, to maximize cluster-wide throughput, these systems intertwine pipeline paths across instances, creating a single point of failure where a single instance interruption propagates to multiple pipelines. Latency prediction approaches~\cite{helix,vidur} rely on extensive profiling, the cost of which becomes intractable as the number of hardware types and configuration combinations increases.
Deploying LLM serving systems on heterogeneous spot GPU clusters introduces challenges that are not addressed by any single line of prior work. Systems designed for spot instances~\cite{spotserve,bamboo-nsdi,parcae-nsdi} assume homogeneous GPU clusters, which are vulnerable to correlated failures where all instances of the same type are simultaneously interrupted. Systems designed for heterogeneous clusters~\cite{thunderserve,helix,hexgen} do not target spot instance environments and therefore lack fault tolerance mechanisms. For instance, to maximize cluster-wide throughput, these systems intertwine pipeline paths across instances, creating a single point of failure where a single instance interruption propagates to multiple pipelines. Latency prediction approaches~\cite{helix,vidur} rely on extensive profiling, the cost of which becomes intractable as the number of hardware types and configuration combinations increases.

The DP-based model placement optimizer of ShuntServe determines the optimal parallelization strategy for each pipeline in the heterogeneous GPU cluster while constraining each instance to a single pipeline, confining the impact of a spot interruption to a single pipeline. The analytical serving performance estimator based on the roofline model, combined with beam search, enables rapid and efficient exploration of the combinatorially explosive search space across a heterogeneous GPU cluster. Output-preserving request migration recovers interrupted requests through a recomputation-based approach, and concurrent initialization via the shared tensor store minimizes downtime by overlapping the initialization of a new pipeline with ongoing serving. These mechanisms enable ShuntServe to effectively handle spot interruptions and maximize the utilization of heterogeneous GPUs for cost-efficient LLM serving.

Table~\ref{tab:related_work_comparison} summarizes the key differences between ShuntServe and existing LLM inference systems. For performance estimation, profiling-based approaches become infeasible in the heterogeneous pipeline setting of ShuntServe and HexGen, where the number of configurations grow combinatorially. Both systems instead rely on an analytical model with one-time per hardware calibration. For optimal model placement, prior systems adopt various strategies such as exhaustive search (i.e., Vidur, SpotServe), a genetic algorithm with DP (i.e., HexGen), and a two-phase optimization combining beam search and DP (i.e., AlpaServe). ShuntServe employs a one-phase optimization that prunes DP entries via beam search, unifying cluster grouping and layer partitioning in a single procedure. Finally, SpotServe handles spot interruptions only in homogeneous clusters, while ShuntServe extends fault tolerance to heterogeneous clusters.

\section{Conclusion}\label{sec:conclusion}
This paper presents ShuntServe, the first cost-efficient distributed LLM serving system designed for heterogeneous spot GPU clusters. ShuntServe achieves 1.42$\times$ higher throughput on Llama-3.1-70B than the state-of-the-art homogeneous serving system and 1.35$\times$ higher throughput on Qwen3-32B than state-of-the-art heterogeneous baselines through the DP-based model placement optimizer with a roofline model-based analytical serving performance estimator. ShuntServe further achieves 31.9\% and 31.2\% cost efficiency improvements over on-demand instances for offline and online serving, respectively, through output-preserving request migration and concurrent initialization via the shared tensor store.

\section*{Acknowledgements}
This work was supported by Institute of Information \& communications Technology Planning \& Evaluation (IITP) grant funded by the Korea government (MSIT) (RS-2022-00144309 \& RS-2025-25441560 \& RS-2026-25492200)

\bibliographystyle{IEEEtran}
\bibliography{ShuntServe}

\end{document}